\documentclass[twocolumn, tighten, numberedappendix, twocolappendix, floatfix]{aastex631}

\usepackage{amsmath,amstext}
\usepackage[utf8]{inputenc}
\usepackage{graphicx}
\usepackage{pgfplots}
\usepackage{tikz-3dplot}
\tdplotsetmaincoords{60}{115}
\pgfplotsset{compat=newest}
\usepackage{graphicx}
\usepackage{dcolumn}
\usepackage{bm}
\usepackage[utf8]{inputenc}
\usepackage[T1]{fontenc}
\usepackage{mathptmx}
\usepackage{amsmath}
\usepackage{mathtools}
\usepackage[none]{hyphenat}
\usepackage{ulem}

\providecommand{\sorthelp}[1]{}

\usepackage{blindtext}

\newcommand{\SkyPercentage}{75\%}

\newcommand{\dg}{$^{\circ}$}

\newcommand{\bs}{boresight}

\usepackage{academicons}
\definecolor{orcidlogocol}{HTML}{A6CE39}
\definecolor{citecolor}{rgb}{0.08,0.30,0.85}
\hypersetup{citecolor=citecolor,filecolor=cyan}

\usepackage{color}
\usepackage{tikz}
\usetikzlibrary{shapes}

\usepackage[encapsulated]{CJK}
\newcommand{\cntext}[1]{\begin{CJK}{UTF8}{gbsn}#1\end{CJK}}

\usepackage{devanagari}

\shorttitle{CLASS $90\,\mathrm{GHz}$ Optical Characterization \& Calibration}
\shortauthors{Datta et al.}

\begin{document}
\newcommand{\markerzero}{\raisebox{0.6pt}{\tikz{\node[draw,scale=0.5,circle,fill=black](){};}}}
\newcommand{\markerone}{\raisebox{0.6pt}{\tikz{\node[draw,scale=0.5,circle,fill=blue](){};}}}
\newcommand{\markertwo}{\raisebox{0.6pt}{\tikz{\node[draw,scale=0.5,circle,fill=orange](){};}}}
\newcommand{\markerthree}{\raisebox{0.7pt}{\tikz{\node[draw,scale=0.4,regular polygon, regular polygon sides=4,fill=blue](){};}}}
\newcommand{\markerfour}{\raisebox{0.7pt}{\tikz{\node[draw,scale=0.4,regular polygon, regular polygon sides=4,fill=orange](){};}}}

\title{Cosmology Large Angular Scale Surveyor (CLASS):\\ $90\,\mathrm{GHz}$ Telescope Pointing, Beam Profile, Window Function, and Polarization Performance}

\author[0000-0003-3853-8757]{Rahul Datta}
\affiliation{Department of Astronomy and Astrophysics, University of Chicago, 5640 South Ellis Avenue, Chicago, IL 60637, USA}
\affiliation{The William H. Miller III Department of Physics and Astronomy, Johns Hopkins University, 3701 San Martin Drive, Baltimore, MD
21218, USA}

\author{Michael~K. Brewer}
\affiliation{The William H. Miller III Department of Physics and Astronomy, Johns Hopkins University, 3701 San Martin Drive, Baltimore, MD
21218, USA}

\author[0000-0002-0552-3754]{Jullianna Denes~Couto}
\affiliation{The William H. Miller III Department of Physics and Astronomy, Johns Hopkins University, 3701 San Martin Drive, Baltimore, MD
21218, USA}

\author[0000-0001-6976-180X]{Joseph Eimer}
\affiliation{The William H. Miller III Department of Physics and Astronomy, Johns Hopkins University, 3701 San Martin Drive, Baltimore, MD
21218, USA}

\author[0000-0002-4820-1122]{Yunyang Li (\cntext{李云炀}\!\!)} 
\affiliation{The William H. Miller III Department of Physics and Astronomy, Johns Hopkins University, 3701 San Martin Drive, Baltimore, MD
21218, USA}

\author[0000-0001-5112-2567]{Zhilei Xu (\cntext{徐智磊}\!\!)}
\affiliation{MIT Kavli Institute, Massachusetts Institute of Technology, 77 Massachusetts Avenue, Cambridge, MA 02139, USA}
\affiliation{Department of Physics and Astronomy, University of Pennsylvania, 209 South 33rd Street, Philadelphia, PA 19104, USA}

\author[0000-0001-7941-9602]{Aamir Ali}
\affiliation{The William H. Miller III Department of Physics and Astronomy, Johns Hopkins University, 3701 San Martin Drive, Baltimore, MD
21218, USA}

\author[0000-0002-8412-630X]{John~W. Appel}
\affiliation{The William H. Miller III Department of Physics and Astronomy, Johns Hopkins University, 3701 San Martin Drive, Baltimore, MD
21218, USA}

\author[0000-0001-8839-7206]{Charles L. Bennett}
\affiliation{The William H. Miller III Department of Physics and Astronomy, Johns Hopkins University, 3701 San Martin Drive, Baltimore, MD
21218, USA}

\author[0000-0001-8468-9391]{Ricardo Bustos}
\affiliation{Departamento de Ingenier\'{i}a El\'{e}ctrica, Universidad Cat\'{o}lica de la Sant\'{i}sima Concepci\'{o}n, Alonso de Ribera 2850, Concepci\'{o}n, Chile}

\author[0000-0003-0016-0533]{David T. Chuss}
\affiliation{Department of Physics, Villanova University, 800 Lancaster Avenue, Villanova, PA 19085, USA}

\author[0000-0002-7271-0525]{Joseph~Cleary}
\affiliation{The William H. Miller III Department of Physics and Astronomy, Johns Hopkins University, 3701 San Martin Drive, Baltimore, MD
21218, USA}

\author[0000-0002-1708-5464]{Sumit Dahal}
\affiliation{NASA Goddard Space Flight Center, 8800 Greenbelt Road, Greenbelt, MD 20771, USA}
\affiliation{The William H. Miller III Department of Physics and Astronomy, Johns Hopkins University, 3701 San Martin Drive, Baltimore, MD
21218, USA}

\author[0000-0002-1052-0339]{Francisco Espinoza}
\affiliation{Departamento de Ingenier\'{i}a El\'{e}ctrica, Universidad Cat\'{o}lica de la Sant\'{i}sima Concepci\'{o}n, Alonso de Ribera 2850, Concepci\'{o}n, Chile}

\author[0000-0002-4782-3851]{Thomas~Essinger-Hileman}
\affiliation{NASA Goddard Space Flight Center, 8800 Greenbelt Road, Greenbelt, MD 20771, USA}

\author{Pedro Flux\'{a}}
\affiliation{Instituto de Astrof\'{i}sica, Pontificia Universidad Cat\'{o}lica de Chile,
Vicu\~{n}a Mackenna 4860, Chile}

\author[0000-0003-1248-9563]{Kathleen Harrington}
\affiliation{High Energy Physics Division, Argonne National Laboratory, 9700 S. Cass Avenue, Argonne, IL, USA 60439}
\affiliation{Department of Astronomy and Astrophysics, University of Chicago, 5640 South Ellis Avenue, Chicago, IL 60637, USA}

\author{Kyle Helson}
\affiliation{Center for Space Sciences and Technology, University of Maryland Baltimore County, Baltimore, MD 21250}
\affiliation{NASA Goddard Space Flight Center, 8800 Greenbelt Road, Greenbelt, MD 20771, USA}
\affiliation{Center for Research and Exploration in Space Science and Technology, NASA/GSFC, Greenbelt, MD 20771}

\author[0000-0001-7466-0317]{Jeffrey Iuliano}
\affiliation{Department of Physics and Astronomy, University of Pennsylvania, 209 South 33rd Street, Philadelphia, PA 19104, USA}
\affiliation{The William H. Miller III Department of Physics and Astronomy, Johns Hopkins University, 3701 San Martin Drive, Baltimore, MD
21218, USA}

\author{John Karakla}
\affiliation{The William H. Miller III Department of Physics and Astronomy, Johns Hopkins University, 3701 San Martin Drive, Baltimore, MD
21218, USA}

\author[0000-0003-4496-6520]{Tobias~A. Marriage}
\affiliation{The William H. Miller III Department of Physics and Astronomy, Johns Hopkins University, 3701 San Martin Drive, Baltimore, MD
21218, USA}

\author{Sasha Novack}
\affiliation{The William H. Miller III Department of Physics and Astronomy, Johns Hopkins University, 3701 San Martin Drive, Baltimore, MD
21218, USA}

\author[0000-0002-5247-2523]{Carolina N\'{u}\~{n}ez}
\affiliation{The William H. Miller III Department of Physics and Astronomy, Johns Hopkins University, 3701 San Martin Drive, Baltimore, MD
21218, USA}

\author[0000-0002-0024-2662]{Ivan L. Padilla}
\affiliation{The William H. Miller III Department of Physics and Astronomy, Johns Hopkins University, 3701 San Martin Drive, Baltimore, MD
21218, USA}

\author[0000-0002-8224-859X]{Lucas Parker}
\affiliation{Space and Remote Sensing, MS D436, Los Alamos National Laboratory, Los Alamos, NM 87544, USA}
\affiliation{The William H. Miller III Department of Physics and Astronomy, Johns Hopkins University, 3701 San Martin Drive, Baltimore, MD
21218, USA}

\author[0000-0002-4436-4215]{Matthew~A.~Petroff}
\affiliation{Center for Astrophysics, Harvard \& Smithsonian, 60 Garden Street, Cambridge, MA 02138, USA}

\author[0000-0001-5704-271X]{Rodrigo Reeves}
\affiliation{CePIA, Departamento de Astronom\'{i}a, Universidad de Concepci\'{o}n, Concepci\'{o}n, Chile}

\author[0000-0003-4189-0700]{Karwan Rostem}
\affiliation{NASA Goddard Space Flight Center, 8800 Greenbelt Road, Greenbelt, MD 20771, USA}

\author[0000-0001-7458-6946]{Rui Shi 
(\cntext{时瑞}\!\!)}
\affiliation{The William H. Miller III Department of Physics and Astronomy, Johns Hopkins University, 3701 San Martin Drive, Baltimore, MD
21218, USA}

\author[0000-0003-3487-2811]{Deniz A. N. Valle}
\affiliation{The William H. Miller III Department of Physics and Astronomy, Johns Hopkins University, 3701 San Martin Drive, Baltimore, MD
21218, USA}

\author[0000-0002-5437-6121]{Duncan J. Watts}
\affiliation{Institute of Theoretical Astrophysics, University of Oslo, P.O. Box 1029 Blindern, N-0315 Oslo, Norway}
\affiliation{The William H. Miller III Department of Physics and Astronomy, Johns Hopkins University, 3701 San Martin Drive, Baltimore, MD
21218, USA}

\author[0000-0003-3017-3474]{Janet L. Weiland}
\affiliation{The William H. Miller III Department of Physics and Astronomy, Johns Hopkins University, 3701 San Martin Drive, Baltimore, MD
21218, USA}

\author[0000-0002-7567-4451]{Edward J. Wollack}
\affiliation{NASA Goddard Space Flight Center, 8800 Greenbelt Road, Greenbelt, MD 20771, USA}

\author[0000-0001-6924-9072]{Lingzhen Zeng}
\affiliation{Center for Astrophysics, Harvard \& Smithsonian, 60 Garden Street, Cambridge, MA 02138, USA}

\correspondingauthor{Rahul Datta}
\email{rahuld@uchicago.edu}


\begin{abstract}
The Cosmology Large Angular Scale Surveyor (CLASS) is a telescope array that observes the cosmic microwave background (CMB) over $\sim \SkyPercentage{}$ of the sky from the Atacama Desert, Chile, at frequency bands centered near 40, 90, 150, and $220\,\mathrm{GHz}$. CLASS measures the large angular scale CMB polarization to constrain the tensor-to-scalar ratio and the optical depth to last scattering. This paper presents the optical characterization of the $90\,\mathrm{GHz}$ telescope. Observations of the Moon establish the pointing while dedicated observations of Jupiter are used for beam calibration. The standard deviations of the pointing error in azimuth, elevation, and boresight angle are 1.3$^{\prime}$, 2.1$^{\prime}$, and 2.0$^{\prime}$, respectively, over the first three years of observations. This corresponds to a pointing uncertainty $\sim 7\%$ of the beam's full width at half maximum (FWHM). The effective azimuthally-symmetrized instrument 1D beam estimated at $90\,\mathrm{GHz}$ has a FWHM of $0.620\pm0.003$\dg{} and a solid angle of $138.7\pm0.6$(stats.)$\pm1.1$(sys.)$\mu{\rm sr}$ integrated to a radius of 4\dg{}. The corresponding beam window function drops to $b_\ell^2 = 0.93,\,0.71,\,0.14$ at $\ell=30,\,100,\,300$, respectively. Far-sidelobes are studied using detector-centered intensity maps of the Moon and measured to be at a level of $10^{-3}$ or below relative to the peak. The polarization angle of Tau A estimated from preliminary survey maps is $149.6\pm0.2$(stats.)\dg{} in equatorial coordinates. Instrumental temperature-to-polarization ($T\rightarrow P$) leakage fraction, inferred from per detector demodulated Jupiter scan data, has a monopole component at the level of $1.7\times 10^{-3}$, a dipole component with an amplitude of $4.3\times 10^{-3}$, and no evidence of quadrupolar leakage.
\end{abstract}

\keywords{\href{http://astrothesaurus.org/uat/799}{Astronomical instrumentation (799)}; \href{http://astrothesaurus.org/uat/322}{Cosmic microwave background radiation (322)}; \href{http://astrothesaurus.org/uat/435}{Early Universe (435)}; \href{http://astrothesaurus.org/uat/1146}{Observational Cosmology (1146)}; \href{http://astrothesaurus.org/uat/1127}{Polarimeters (1127)}}


\vskip 5.8mm plus 1mm minus 1mm
\vskip1sp
\section{Introduction}
\vskip1sp
Over the last few decades, measurements of the cosmic microwave background (CMB) temperature anisotropy have provided the most precise picture of the early Universe and validated the $\Lambda$CDM model of cosmology~\citep[e.g.,][]{1996COBE_anisotropy,benn13,Hinshaw13,2020A&A...641A...5P}. The polarization of the CMB provides a unique window into the physics of inflation~\citep{guth81,albr82,lind82,2009AIPC.1141...10B}, a sensitive probe of the optical depth to reionization \citep{zald1997}, and a means to measure the neutrino mass sum through its lensing~\citep{2009AIPC.1141..121S,Neutrino_Allison}. It also enables a multitude of other astrophysical studies through its cross-correlation with surveys at other wavelengths~\citep[e.g.,][]{2016MNRAS.459...21K,2019JCAP...10..057F,2019A&A...625L...4H,Namikawa_2019,2021PhRvD.103f3513S,2021MNRAS.500.2250D}. The decomposition of CMB polarization into even-parity E and odd-parity B modes is of special cosmological significance. While cosmological (i.e., not foregound) B modes are only produced by inflationary gravitational waves sourced by tensor perturbations (``primordial B modes'') or conversion of E to B modes through gravitational lensing (``lensing B modes'') \citep{kami97,zald97}, E modes can be produced by scalar as well as tensor perturbations. Continuously improving measurements of the E-mode polarization have further supported the standard model~\citep[e.g.,][]{dasi,readhead04,Hinshaw13,henn18,kusa18,2020A&A...641A...5P,2020JCAP...12..045C,2021ApJ...908..199R}. Detections of the lensing B modes which peak at multipole $\ell \approx 1000$ have been reported by several experiments~\citep[e.g.,][]{pola14,das14,keis15,bice16,loui17,pola17,omor17,keck18,2020A&A...641A...8P}. Measurements targeting the primordial B modes expected to peak at degree scales have been progressing in recent years~\citep{2020PhRvD.101l2003S,2021PhRvL.127o1301A,2022ApJ...931..101A}, improving the upper limit on the tensor-to-scalar ratio $r$. Major ground-based CMB surveys, either just starting or being planned as of this writing, aim to precisely measure the E and B modes at $\ell>30$ enabled by increased sensitivity, control over systematics, and modeling of foregrounds~\citep[][]{BICEPArrayOverview18,simons19whitepaper, 2019arXiv190704473A}. 

\begin{figure*}
    \centering
    \includegraphics[width=0.74\linewidth]{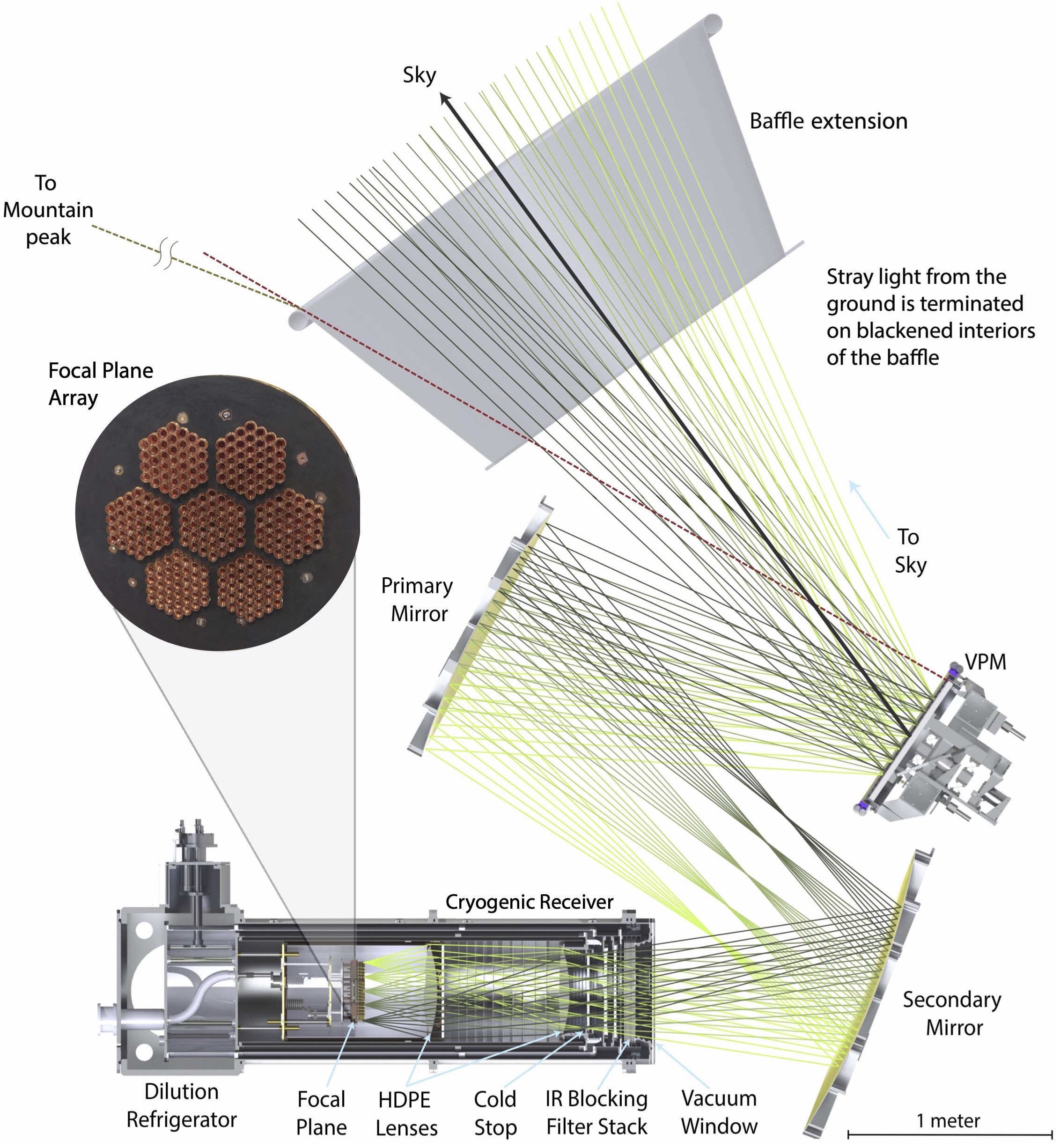}
    \caption{\textbf{Ray-trace schematic of the 90$\,\mathrm{\textbf{GHz}}$ telescope showing the major optical components}. Color-coded rays tracing the light's path originate from five fields on the sky and converge at five feedhorns in the focal plane. The VPM is the first optical element in the path of the incoming light through a co-moving baffle. The co-moving baffle comprises a ``cage'' structure (shown in Figure~\ref{fig:bs_rotation}, which houses the fore-optics and the receiver), and an ``extension.'' The solid black line shows the boresight pointing, the dashed red line connects the top of the VPM mirror to the rolled top of the extension. At 45\dg{} elevation and 0\dg{} rotation about boresight, the red line (when extended) clears the highest ground feature by 6\dg{} in elevation. Stray light from the ground entering through other sides of the baffle is terminated on absorber covered interiors of the baffle (see Section~\ref{subsec:far_sidelobe} for further discussion). The primary and secondary mirrors produce an image of the VPM near the cold stop. High density polyethylene (HDPE) lenses then focus the light onto  feedhorn-coupled dual-polarization detectors with speed $f/1.8$. Seven detector modules, each comprising 37 pixels, constitute the focal plane array. A picture of the installed detector modules is shown, surrounded by an absorber coated copper structure serving as a field stop. See Appendix A for further details of the optical design. }
    \label{fig:ray_schematic}
\end{figure*}

The Cosmology Large Angular Scale Surveyor (CLASS)~\citep{essi14,harr16} is designed to measure the polarization of the CMB over a range of large angular scales ($\ell \lesssim 200$) sensitive to both the recombination and reionization~\citep{kami16} signatures in the B-mode signal associated with primordial gravitational waves from inflation. Observing $\sim \SkyPercentage{}$ of the sky in four frequency bands centered near 40, 90, 150, and $220\,\mathrm{GHz}$ from a high altitude site in the Atacama Desert of Chile, CLASS uses rapid front-end polarization modulation~\citep{harr18} and a scanning strategy enabling adequate cross-linking to recover the polarization signal at large angular scales on the sky~\citep{mill16}. The large angular scale E-mode measurement will improve constraints on the optical depth to reionization~\citep{2018_tau_Duncan}, complementing 21~cm measurements~\citep{2018Natur.555...67B, 2022NatAs...6..607S}, and informing studies of high-redshift galaxy formation. The improved constraints on the optical depth $\tau$ to reionization will improve constraints on parameters degenerate with $\tau$, such as the sum of neutrino masses~\citep{Neutrino_Allison}. The $40\,\mathrm{GHz}$ telescope~\citep{2014_40GHz_Detector_John,appe19} has been observing since 2016, while the $90\,\mathrm{GHz}$ telescope achieved first light in 2018~\citep{daha18}. The $90\,\mathrm{GHz}$ focal plane was upgraded in the austral winter of 2022~\citep{10.1117/12.2630081}. A third telescope housing a dichroic receiver sensitive to both 150 and $220\,\mathrm{GHz}$ frequency bands \citep{daha19} began observations in 2019. A fourth telescope at $90\,\mathrm{GHz}$ is planned. Other ground-based, balloon-borne, and space-based experiments seeking to target CMB polarization on the largest angular scales ($\ell<30$) include  Taurus \citep{taurus20apra}, Groundbird \citep{2020JLTP..200..384L},  LSPE \citep{2021JCAP...08..008A},  LiteBIRD \citep{litebird22}, and QUIJOTE \citep{quijote23}. 

When mapping the CMB, the measured signal is the true sky signal convolved with the telescope beam pattern. Therefore, extensive characterization of the telescope's optical response~\cite[e.g.,][]{page03,hass13,2014A&A...571A...7P,pan18,bice19} is critical to recovering the true signal as well as understanding and quantifying the impact of instrumental systematics. The absolute telescope pointing and its stability over the observing season, beam profile and corresponding window function, detector polarization angles, and temperature-to-polarization leakage are particularly important in this regard. 

This paper presents the optical characterization of the CLASS $90\,\mathrm{GHz}$ telescope during its operation from July 2018 to May 2022, focusing on pointing and beam calibration, beam window function, detector polarization angles, far-sidelobes, and temperature-to-polarization leakage. We refer to this period of the survey as ``Era~2". This work builds on the preliminary pointing stability and beam measurements at 90, 150, and $220\,\mathrm{GHz}$ presented in~\cite{10.1117/12.2630649}. The optical characterization of the CLASS $40\,\mathrm{GHz}$ telescope during the first two years of its operation (``Era~1") was described by~\cite{2020ApJ...891..134X}. Other Era~1 papers address telescope calibration, efficiency, and sensitivity~\citep{appe19,2022ApJS..262...52A}; circular polarization~\citep{Padilla2019, Petroff2019}; and polarization modulation and long-timescale instrument stability~\citep{harr21}. The CLASS data reduction pipeline and polarization maps at 40~GHz are presented in~\cite{Li23}; angular power spectra and map-based results are presented in \cite{Eimer23}. Measurements of the disk-averaged absolute Venus brightness temperature using CLASS have been reported in~\cite{Dahal_2021} and \cite{Dahal_2023}. While similar to Era~1, the optical performance of the $40\,\mathrm{GHz}$ telescope during Era~2 is summarized in~\cite{2022ApJ...926...33D}, which presents on-sky performance of all three deployed CLASS instruments observing in four frequency bands between 30 and $250\,\mathrm{GHz}$. Long-timescale stability of the time-stream data across all CLASS frequencies is presented in~\cite{10.1117/12.2629723}. Optical characterization of the CLASS $150/220\,\mathrm{GHz}$ telescope will be presented in a future paper. 

This paper is organized as follows. The CLASS $90\,\mathrm{GHz}$ instrument and survey scan strategy, including telescope configuration, observing modes, and data used in this analysis are described in Section~\ref{sec:instrument&data}. The telescope pointing calibration and its stability are presented in Section~\ref{sec:pointing_analysis}. In Section~\ref{sec:beam}, measurements of the effective beam in intensity made from dedicated planet scans and the beam window function for CMB analysis are described. Constraints on crosstalk and far-sidelobes are discussed. Polarization angle calibration with Tau~A is presented in Section~\ref{sec:pol_ang}. Constraints on instrumental temperature-to-polarization leakage are discussed in Section~\ref{sec:TtoP_leakage}. Finally, the impacts of some beam systematic effects are discussed in Section~\ref{sec:systematics}, and we summarize in Section~\ref{sec:summary}. The pointing model and the beam window function are products of this work that are directly used for cosmological analysis.

\vskip 5.8mm plus 1mm minus 1mm
\section{Instrument \& Observations}
\label{sec:instrument&data}

\begin{figure*}
    \centering
    \includegraphics[width=1\linewidth]{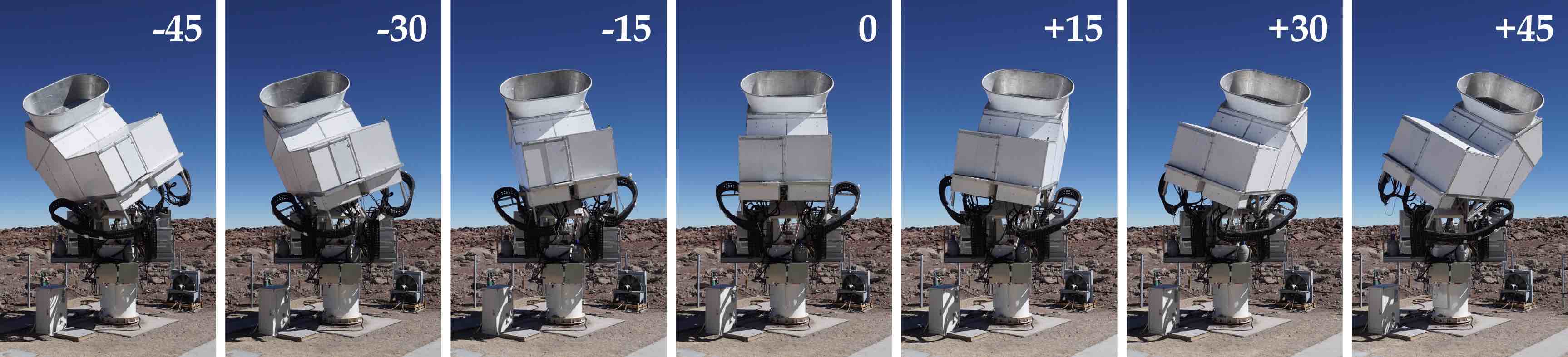}
    \caption{Photos of a telescope mount showing the seven boresight rotation angles. Each of two mounts can support two telescopes. This rotation keeps the telescope \bs{} pointing unchanged while rotating the detector antenna orientation on the sky in steps of 15\dg{} within a 90\dg{} range, thus enabling measurement of the polarization signal projected onto different orientations.}
    \label{fig:bs_rotation}
\end{figure*}

A schematic of the $90\,\mathrm{GHz}$ CLASS telescope is shown in Figure~\ref{fig:ray_schematic} with corresponding ray-trace and main components rendered. The optical design is similar to the design of the $40\,\mathrm{GHz}$ telescope~\citep{eime12}. The first optical element in the path of the sky signal is a fixed-wire grid in front of and parallel to a movable-mirror with a 60~cm clear aperture, which is referred to as a variable-delay polarization modulator (VPM) \citep{chus12, harr18}. The VPM modulates the polarized signal at 10~Hz before any potential polarization signals are introduced by the telescope. This front-end modulation to frequencies higher than atmospheric and instrumental drifts enables recovery of the largest angular scale modes while suppressing temperature-to-polarization leakage from atmospheric signals. It is key to achieving the science goals of CLASS. The fore-optics, comprising the primary and secondary mirrors, produce an image of the cold stop near the VPM. An ultra-high-molecular-weight polyethylene (UHMWPE) vacuum window of thickness $3.175\,\mathrm{mm}$ on the cryogenic receiver followed by a stack of infrared (IR)-blocking filters~\citep{essi14, 2018Cryostat_Jeff} allow a high degree of microwave transmission while strongly suppressing infrared radiation. The vacuum window is anti-reflection (AR) coated on both sides with a layer of porous
polytetrafluoroethylene (PTFE) whose thickness is optimized for maximum in-band transmission. Two high-density polyethylene lenses re-image the light from each field (See Figure~\ref{fig:ray_schematic}) onto the focal plane with focal ratio $\sim1.8$ of the lens closest to the focal plane. The VPM is significantly under-illuminated with the beam tapering down below $-30\,\mathrm{dB}$ at the edge of the VPM. This minimizes edge effects that could produce unwanted sidelobes. For details of the optical design, see Appendix A.

The focal plane consists of seven detector modules. Each module has 37 smooth-walled feedhorns~\citep{5398898,daha18} that couple the incoming light to microfabricated antenna probes on a $100\,\mathrm{mm}$ planar detector wafer~\citep{Rostem:2016}. The feeds are hexagonal close packed and spaced by approximately $14\,\mathrm{mm}$, which is $\sim2.3\cdot F\lambda$, where $\lambda$ is the free space wavelength at $90\,\mathrm{GHz}$, and $F\sim1.8$ is the effective focal ratio. The feeds illuminate the edge of the cold stop at $-8.4\,\mathrm{dB}$, resulting in high spill efficiency and low levels of unwanted diffraction as the beam propagates through the telescope. The $90\,\mathrm{GHz}$ telescope's field of view (FOV) is  approximately $23^\circ$ in diameter. Each pixel, fed by a feedhorn~\citep{doi:10.1063/5.0049526}, is sensitive to both orthogonal states of linear polarization enabled by an orthomode transducer (OMT) on the detector wafer. Microwave transmission lines pick up the signal from the OMT antenna probes and route them to transition-edge sensor (TES) bolometers. Each pixel on the focal plane has two TESs for measuring the power in both polarization states. In addition to these, there are also TES bolometers on each wafer that are not electromagnetically coupled to antennas. These unilluminated detectors are
useful for tracking focal plane temperature variations and other possible
sources of systematic error.
The focal plane is cooled to $\sim50\,\mathrm{mK}$ by a dilution refrigerator \citep{2018Cryostat_Jeff}. The detectors are read out through a time-division-multiplexed (TDM) scheme~\citep{daha18} utilizing superconducting quantum interference devices~\citep[SQUID,][]{reintsema2003} and ambient temperature multi-channel electronics~\citep[MCE,][]{ubc_mce}.  For the remainder of the paper, we will refer to optically-coupled detectors as ``optical detectors'' (or simply as ``detectors''), whereas the isolated bolometers will be referred to as ``dark detectors''. SQUIDs that are not connected to
the TES bolometers will be referred to as ``dark SQUIDs''. The optical detectors are furthermore distinguished by their polarization-sensitive orientation on the sky. A ``$-45$\dg{} detector'' has an on-sky polarization direction rotated approximately 45\dg{} clockwise from vertical in receiver coordinates viewing the sky from the telescope. Similarly a ``$+45$\dg{} detector'' has its polarization rotated 45\dg{} counter-clockwise from vertical~\citep[the choice of $\pm45^\circ$ detector polarization angles is required by the VPM,][]{harr21}. The instrument passband spans $77-108\,\mathrm{GHz}$, centered at $\sim 91\,\mathrm{GHz}$~\citep{daha18}. The on-sky performance of the $90\,\mathrm{GHz}$ detectors is presented in~\cite{2022ApJ...926...33D}.

\begin{figure*}
    \centering
    \includegraphics[width=0.95\linewidth]{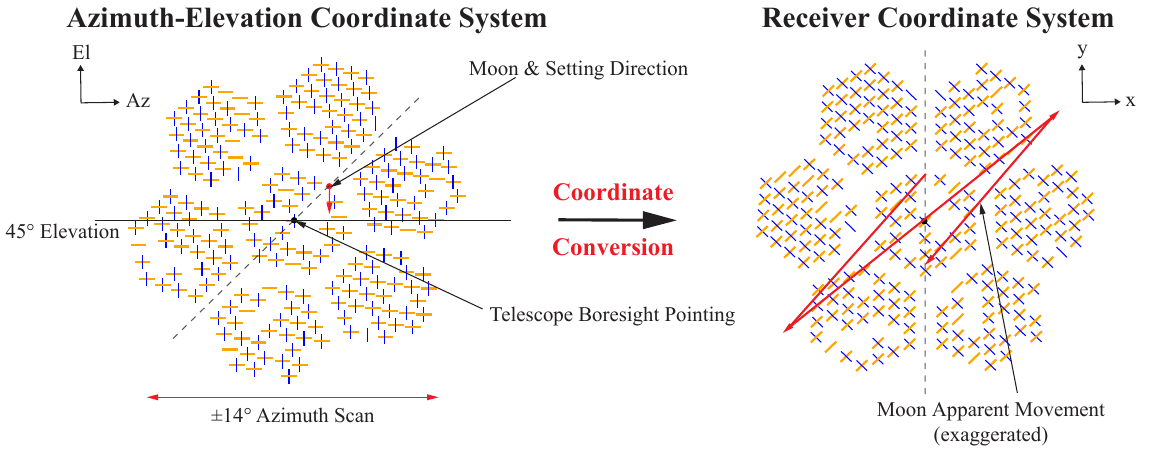}
    \caption{\textbf{Moon scan illustration and coordinate-system conversion}. {\textit{Left}}: Schematic of the focal plane detector array, centered at a scan elevation of $45$\dg{} with a $-45$\dg{} \bs{} rotation. Each blue (orange) bar represents a detector sensitive to $-45$\dg{} ($+45$\dg{}) polarization orientation. The Moon slowly rises or sets (the red dot indicates the position of the Moon for an example scan and the red arrow indicates that it is setting in this example) as the telescope scans $\pm$14\dg{} along the azimuthal direction. In the azimuth-elevation coordinate system, the focal plane array spans 23.6\dg{} in azimuth and 24\dg{} in elevation. The dashed line represents the plane of optical symmetry for the telescope. {\textit{Right}}: Moon positions are transformed to the ``Receiver Coordinate System'', where the telescope \bs{} pointing is the origin. Every detector has fixed $\Delta x$ and $\Delta y$ angular offsets while the Moon appears to zigzag across the array. In the azimuth-elevation coordinate system, both the telescope and the Moon are moving, while in the receiver coordinate system only the Moon is moving. The apparent movement of the Moon across the array due to Earth's rotation during one such scan is indicated, where the spacing of the zigzag path is exaggerated for illustration. }
    \label{fig:sky_tele_conv}
\end{figure*}


A three-axis mount, shown in Figure \ref{fig:bs_rotation}, points the $90\,\mathrm{GHz}$ telescope along with the telescope at $40\,\mathrm{GHz}$. A second identical mount supports a third telescope that houses a dichroic receiver observing at 150 and $220\,\mathrm{GHz}$ simultaneously and will house a second $90\,\mathrm{GHz}$ telescope. The two mounts can rotate in azimuth, elevation, and \bs{} independent of each other. During CMB survey observations, the mounts nominally rotate 720\dg{} in azimuth at a fixed elevation of 45\dg{}, before reversing around and scanning in the opposite direction, thus scanning the sky in large circles. During the majority of Era~2 CMB observations, the telescope scanned in azimuth at 2\dg{}$\mathrm{s}^{-1}$ at a constant elevation. For the first few months of Era~2, the telescope scanned at 1\dg{}$\mathrm{s}^{-1}$. The faster scanning mode is part of a suite of improvements implemented in Era~2 aimed at improving long time-scale stability. During the day, the telescope boresight is pointed at least 20\dg{} away from the Sun, increasing the frequency of the turnarounds and reducing the azimuthal range from the nominal 720\dg{}. The telescopes observe $\sim \SkyPercentage{}$ of the sky every day. Complete measurement of Stokes $Q$ and $U$ is enabled by the Earth's spin and daily rotations of the telescope \bs{} platform, which rotate the orientation of the detector polarization on the sky. The rotation about the \bs{} spans 90\dg{} as depicted in Figure~\ref{fig:bs_rotation}, cycling through seven angles ($-45$\dg{}, $-30$\dg{}, $-15$\dg{}, 0\dg{}, $+15$\dg{}, $+30$\dg{}, $+45$\dg{}) each week. This scan strategy enables cross-linking on different time scales with each pixel on the sky being observed with different scanning directions and detector orientations.

\vskip1sp

Calibration campaigns are periodically undertaken by performing dedicated scans of bright sources, namely, the Moon, Jupiter, and Venus. Moon scans are used for pointing calibration, whereas planet scans are used for beam characterization since the bright signal from the Moon nearly saturates the detectors. The planets are essentially point sources given the instrument beam size. During the calibration runs, the telescope scans across the source azimuthally, maintaining a constant elevation. Additionally, \bs{} rotations  help improve the sampling of the spatial beam and provide information about the offset of the center of the array which would otherwise be degenerate with the boresight pointing. Given the telescope FOV, the azimuthal scans span $\pm$14\dg{} on the sky, centered on the source, so that beams at the edge of the FOV are measured to adequately large angles $\gtrsim 2$\dg{}. Comparing the observed planet signal with its expected brightness temperature yields a $90\,\mathrm{GHz}$ telescope net end-to-end optical efficiency of 0.4~\citep{2022ApJ...926...33D}. For the remainder of the paper, a ``scan'' in the context of pointing calibration and beam characterization refers to a set of continuous observations across a source as it rises or sets through the array.

\begin{figure*}[!htb]
    \centering
    \includegraphics[width=1\textwidth]{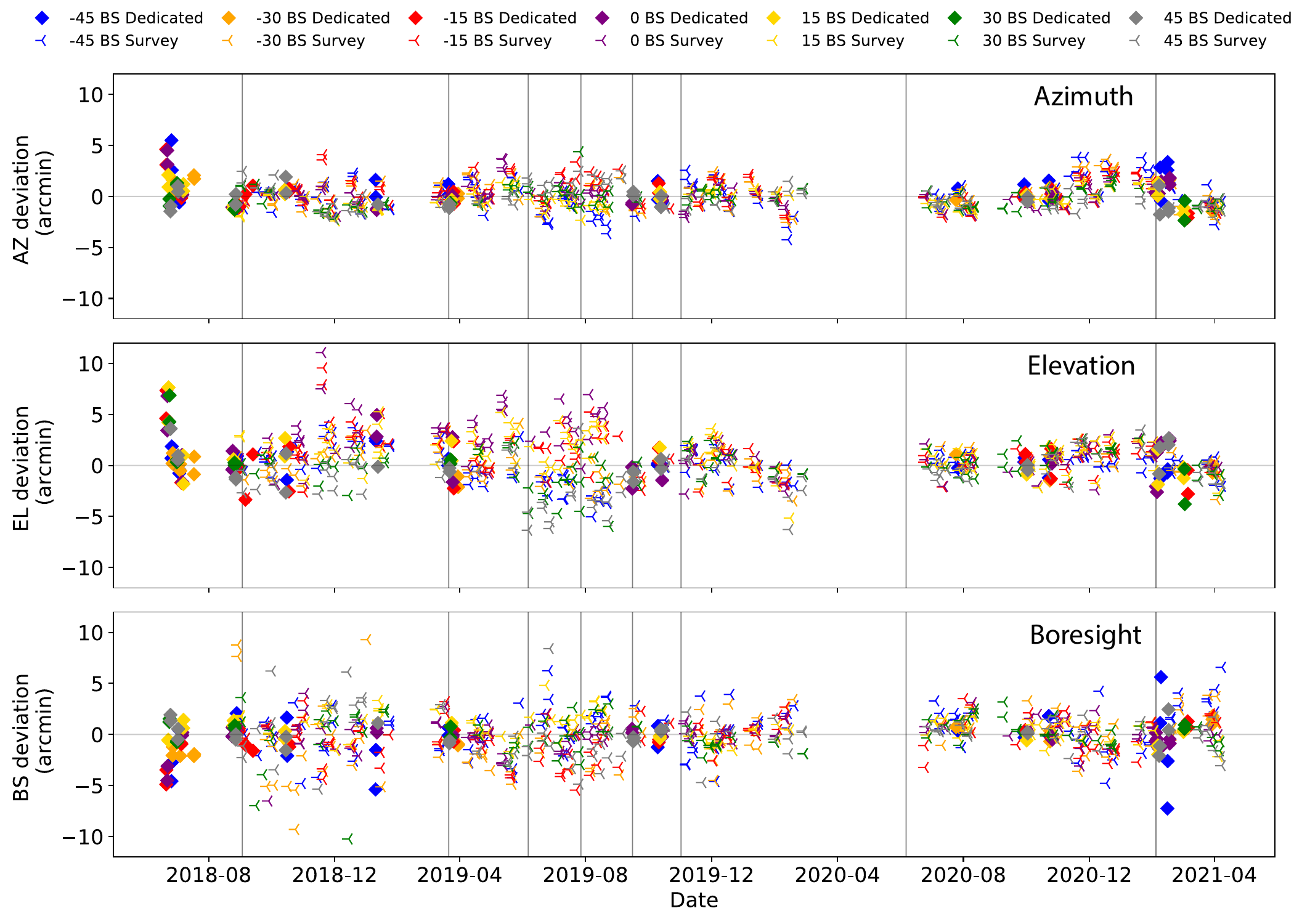}
    \caption{\textbf{Deviations from the pointing model in arcminutes for telescope azimuth, elevation, and boresight rotation angle  inferred from Moon scans over the period of July 2018 through April 2021.} The seven telescope boresight angles ranging from $-45$\dg{} to $+45$\dg{} are indicated by different colors. Diamonds represent deviations inferred from dedicated Moon scans, and tri-left markers from survey Moon scans. Deviations are inferred separately from scans conducted during the Moon rising and setting, when available. Gray vertical lines (eight in total) highlight the start date of each new pointing correction model based on a series of dedicated Moon scans. The standard deviations of the pointing error are $1.3'$, $2.1'$, and $2.0'$ in azimuth, elevation, and boresight rotation angle, respectively, which amount to a combined pointing uncertainty that is $\sim 7\%$ of the beam FWHM. }
    \label{fig:pointing}
\end{figure*}

\vskip 5.8mm plus 1mm minus 1mm
\section{Pointing}
\label{sec:pointing_analysis}
Targeted observations of the Moon are used for pointing calibration. As described by \cite{2020ApJ...891..134X}, during these dedicated Moon scans, the telescope tracks the rising and setting of the Moon in a range of $\pm$14\dg{} as it passes through the focal plane's FOV. Such observations are performed a few times over the observing season, at elevations of 31\dg{}, 45\dg{}, and 60\dg{}. The 31\dg{} and 45\dg{} scans are performed at 1.5\dg{}$\mathrm{s}^{-1}$, while the 60\dg{} scans are performed at 2\dg{}$\mathrm{s}^{-1}$ due to the wider range in azimuthal angle required to attain $\pm$14\dg{} on the sky at this elevation. The full range of boresight angles is covered at each elevation. Following each of these calibration campaigns, the pointing model is updated. 

The pointing analysis begins by calibrating~\citep{2022ApJS..262...52A}, filtering, and downsampling the raw time-ordered data (TOD) from each dedicated scan. Moon-centered intensity maps are generated for each detector, and fit with two-dimensional Gaussian profiles to recover the Moon position. These are used to generate a complete set of per-detector pointing offsets in the azimuth-elevation coordinate system (as a function of azimuth, elevation, and boresight angle), and in the receiver coordinate system (through a spherical coordinate system transformation -- see Figure~\ref{fig:sky_tele_conv}), referred to as the pointing model. 

\vskip1sp

\subsection{Pointing stability}
\label{subsec:pointing_data}
In addition to the dedicated Moon scans, the Moon is observed during CMB survey observations; we refer to these as survey Moon scans. A total of 798 Moon scans conducted over the period of July 2018 through April 2021 were considered for updating pointing models and tracking pointing deviations from the model. Table \ref{tb:pointing} shows the number of scans at each telescope boresight angle. Of these, 142 were dedicated scans and 656 were survey scans. Each selected survey scan included at least 100 detectors scanning the Moon. The survey scans are only used for tracking pointing deviations, because the Moon sampling is sparse in such scans.

\begin{figure*}
\includegraphics[width=0.95\linewidth]{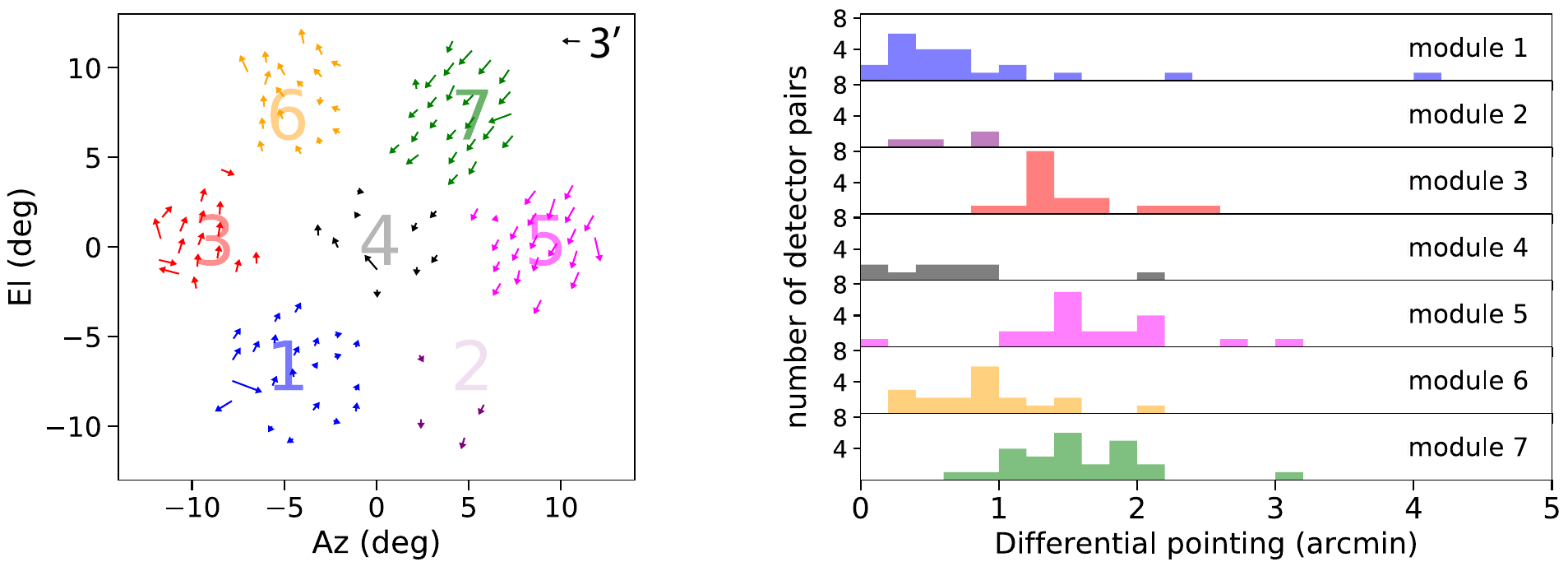}
\caption{\textit{Left:} Measured detector pointing offsets within a co-located pair are represented by vectors originating at the pointing of the $-45$\dg{} oriented detector and directed towards the pointing of the $+45$\dg{} oriented detector on the sky. The length of the vectors are exaggerated by a factor of 20 for visibility (see annotation in the top right corner). The different colors represent the seven modules (module numbers are annotated) that constitute the focal plane. The modules are numbered left to right row-wise starting from the bottom row. \textit{Right:} Histogram showing the distribution of the differential pointing offsets grouped by focal plane module. Most of the differential pointings are within $3'$.}
\label{fig:diff_pointing}
\end{figure*}

For tracking pointing deviations with the data from a dedicated or a survey Moon scan, Moon-centered intensity maps are generated for each detector as before. The Moon position is transformed from azimuth-elevation to the receiver coordinate system, where pointing offsets for each detector are fitted as two-dimensional displacements $\Delta x$, $\Delta y$ with respect to the last updated pointing model. The resulting per detector offsets are projected back on to the sky at the original azimuth, elevation, and boresight rotation angle. These are then used to compute array averaged angular ``on sky'' displacements through a combined minimization of all detector offsets with respect to the expected ``on sky'' detector locations. These constitute the pointing deviations plotted as a function of time in Figure~\ref{fig:pointing}. 

The mean (standard deviations) of these pointing errors derived from dedicated Moon scans only are $0.15'\,(1.3')$ in azimuth, $0.49'\,(1.9')$ in elevation, and $-0.2'\,(1.5')$ in boresight rotation angle. When including survey scans, the pointing errors in azimuth, elevation, and boresight rotation angle are $0.08'\,(1.3')$, $0.51'\,(2.1')$, and $0.01'\,(2.0')$, respectively. The combined pointing uncertainty is $2.5'$, which is $\sim 7\%$ of the beam FWHM. Convolving the measured beam with this measured pointing uncertainty results in an effective broadening of the beam in the survey maps of $\sim 3\%$. This is accounted for in the cosmological data analysis.

\subsection{Differential Pointing}
\label{subsec:pointing_diff}

The pointing of each detector on the sky is well measured, with uncertainties within $10^{\prime\prime}$. Two co-located detectors sharing a feed have sensitivity to orthogonal states of linear polarization. When these point at slightly different locations on the sky, the resulting differential pointing can lead to non-zero polarization systematics in a pair-differencing experiment~\citep{2008PhRvD..77h3003S,2009PhRvD..79f3008M,Fluxa20}. Figure~\ref{fig:diff_pointing} shows the differential pointing between co-located detectors for 115 feeds across the focal plane which have both detectors operational. Module-dependent trends are seen in the differential pointing vectors. The differentials are smallest for the central module and in general larger closer to the edges of the focal plane. The mean differential pointing is $1.2^{\prime}$ with a scatter of  $0.7^{\prime}$ across the focal plane. This is $\sim 3\%$ of the beam FWHM. However, the instrument is designed such that pair-differencing is not required for cosmological analysis. Hence, this differential pointing will not impact any cosmological analysis as long as the pointing of individual detectors on the sky is well measured. 

\begin{deluxetable}{cccc}
\tablenum{1}
\tabletypesize{\small}
\tablecaption{Number of Moon scans at each boresight rotation angle.}
\label{tb:pointing}
\tablehead{
\colhead{Boresight rotation angle} & \colhead{Dedicated} & \colhead{Survey} & \colhead{Percentage}
}
\startdata
$-45$\dg{} & 25 & 101 & 15.7 \\
$-30$\dg{} & 14 & 95 & 13.7 \\
$-15$\dg{} & 22 & 99 & 15.2 \\
$0$\dg{} & 26 & 95 & 15.2 \\
$+15$\dg{} & 20 & 91 & 13.9 \\
$+30$\dg{} & 12 & 88 & 12.5 \\
$+45$\dg{} & 23 & 87 & 13.8 \\
\hline
Total & 142 & 656 & 100 \\
\enddata
\end{deluxetable}

\begin{figure*}
    \centering
    \includegraphics[width=0.99\linewidth]{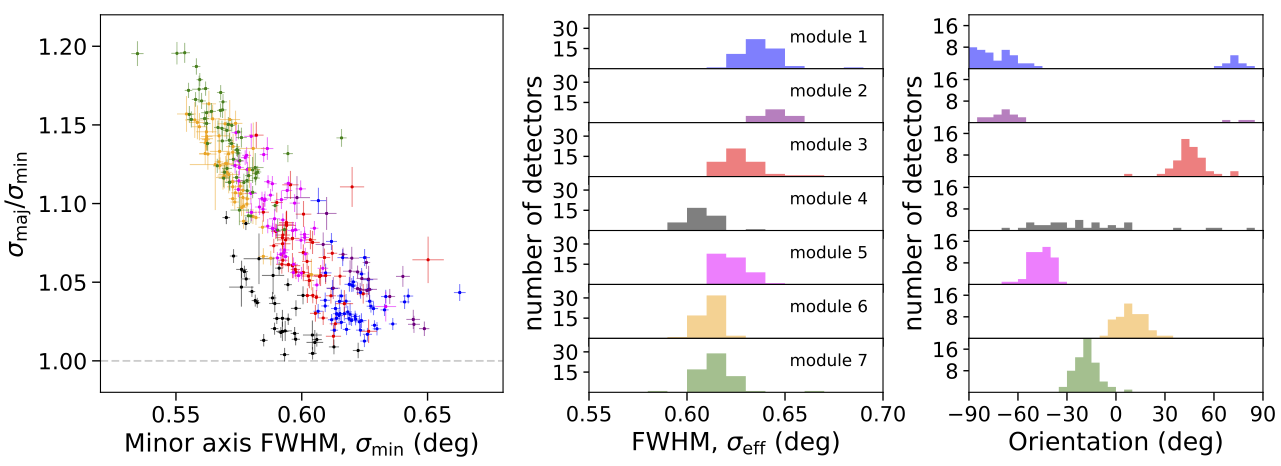}
    \caption{{\textit{Left:}} Ratio of measured FWHM along the major and minor axes plotted against FWHM along the minor axis for each detector. The error bars represent 1$\sigma$ measurement uncertainties. Data points closer to the horizontal dashed line correspond to more circular beams. {\textit{Middle:}} Histogram showing the distribution of the measured beam effective FWHM (Equation~\ref{equ:effective_fwhm}) for all detectors. {\textit{Right:}} Histogram showing the distribution of the beam orientations, defined as the angle on the sky from the local meridian to the beam major axis with positive (negative) corresponding to a counter clockwise (clockwise) direction. In the middle and right panels, each row represents a focal plane module.  The large spread in the distribution for the central module 4 is indicative of nearly circular beams. Boresight rotations help with mitigating the impact of deviations from perfectly symmetric circular beams. The correspondence between colors and modules is the same in all plots.}
    \label{fig:beam_widths}
\end{figure*}

\vskip 5.8mm plus 1mm minus 1mm
\section{Beams}
\label{sec:beam}
For accurate measurements of the CMB angular power spectrum at low $\ell$, it is essential to characterize the beam to large angles. Further, for approximately Gaussian beams with some ellipticity, mismatch in beam shapes and gains between two polarization-sensitive detectors sharing a feed, if not accounted for properly, can lead to spurious polarization signals in a pair-differencing experiment. While we use pair-differencing to probe certain instrumental systematics, the cosmological analysis pipeline works with individual detectors, which mitigates such spurious signals. In this section, measurements of the main-beam parameters are presented and an effective instrument beam is constructed. From this, a radial beam profile for the $90\,\mathrm{GHz}$ instrument is constructed and a beam window function is computed. 

\begin{figure}
    \centering
    \includegraphics[width=1\linewidth]{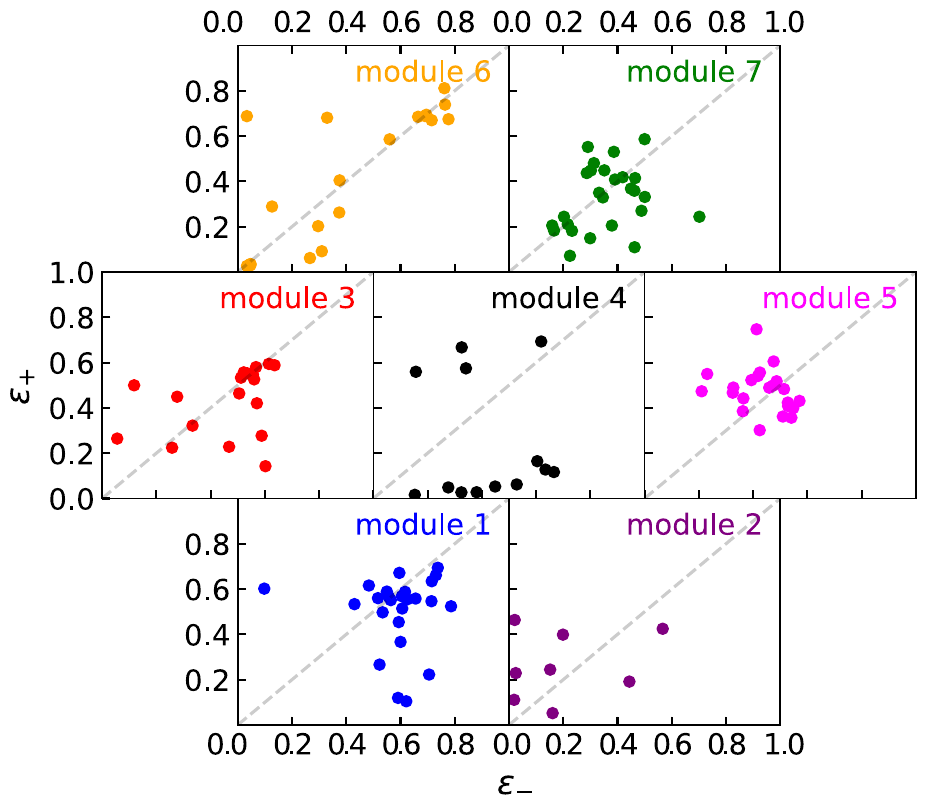}
    \caption{Gain calibration factors for $-45^\circ{}$ oriented detectors ($\epsilon_\mathrm{-}$) plotted against corresponding $+45^\circ{}$ oriented detectors ($\epsilon_\mathrm{+}$) in the pair. Each subplot represents a focal plane module, where the module numbers are annotated in the top right corners. The large scatter in the data and deviations from the diagonal dashed line are due to variations in detector optical efficiencies within each module and within each pair, respectively. Taking into account various detector performance parameters, the wafers from modules 2, 4, 5, and 6 were swapped with new wafers having more uniform optical efficiencies during the $90\,\mathrm{GHz}$ instrument upgrade~\citep{10086549} in 2022.}
    \label{fig:diff_gain}
\end{figure}

\vskip 5.8mm plus 1mm minus 1mm
\subsection{Beam Measurements}
\label{sec:beam_meas}
Dedicated scans of Venus and Jupiter are used to measure the angular response of the detectors on the sky. Such scans were performed soon after the $90\,\mathrm{GHz}$ instrument was deployed in 2018 and then again in 2020, around the time of the Earth's closest approach to Jupiter in the months of June and July. The weather conditions at the site were significantly better during the 2020 Jupiter scans compared to the 2018 scans of Venus and Jupiter. Hence, 70 Jupiter scans conducted in 2020, which provided the highest signal-to-noise measurements, are used for beam characterization. 

Each scan spanned approximately two hours and consisted of multiple passes over the planet at about 2 per minute. The detectors were tuned prior to each scan. A total of 319 of the 407 operational optical detectors in the field~\citep[see][]{2022ApJ...926...33D} detected the signal from Jupiter with signal-to-noise ratio (S/N) greater than 20. For each detector and its corresponding pointing offset, we defined a coordinate system centered at the detector, with the $y$-axis pointing along the local meridian in the receiver coordinate system. 

The data processing for beam measurement is performed in two steps. During the first step, for each scan, the TOD for each detector is read and fit for beam peak amplitude and position in the receiver coordinate system. The root-mean-square (RMS) noise is calculated from data well away from the peak. Data for which the fitted beamwidths fall outside upper or lower bounds or the fitted amplitude is less than 3 times the RMS noise are rejected. Accepted data are cut to 10\dg{} radius centered on the fitted position and recorded along with the fitted position and amplitude.

In the second step, the 10\dg{} radius maps for each detector per scan saved from the first step are read in. Maps with fitted positions more than 0.1\dg{} offset from the expected position are rejected. A parabolic baseline is removed from each pass over the map using data outside a radius of 1.5\dg{} from map center. This radius was chosen to be just greater than the angular distance of the first sidelobe from the beam peak. The maps are then cut to a radius of 4\dg{} and a second linear baseline is removed from each pass using data outside a radius of 1.7\dg{} from map center. This radius is slightly larger due to the smaller range of the 4\dg{} maps. This process primarily removes signal attributable to the varying optical depth of the atmosphere with elevation and low frequency variations in the azimuth scanning direction. The maps are then scaled to a constant fiducial angular diameter of Jupiter taking into account its oblateness~\citep{2011ApJS..192...19W}. The maps are also scaled for atmospheric transmission based on the precipitable water vapor (PWV) recorded at the time of observation and detector zenith angle using the atmospheric model of \cite{982447}. All maps for each detector are evaluated on criteria of S/N and RMS noise. The RMS noise is computed over a radial annulus of 2.25\dg{}$\le \theta \le $4\dg{}. The S/N is defined as the ratio of the fitted amplitude from the first step to the above obtained RMS. For each detector, a median scaled amplitude and median RMS are computed from all maps for the corresponding detector. Maps with RMS greater than 3 times the median RMS or scaled amplitude greater than twice or less than half of the median are rejected. The remaining maps are then averaged weighted by their (S/N)$^{2}$ and fitted to a 2D Gaussian to yield
the measured beam parameters for each detector. (S/N)$^{2}$ weighting was chosen over inverse variance weighting due to its immunity to calibration error, which tends to overweight the scans at the low end of the calibration range.

\begin{figure*}
\includegraphics[width=0.95\linewidth]{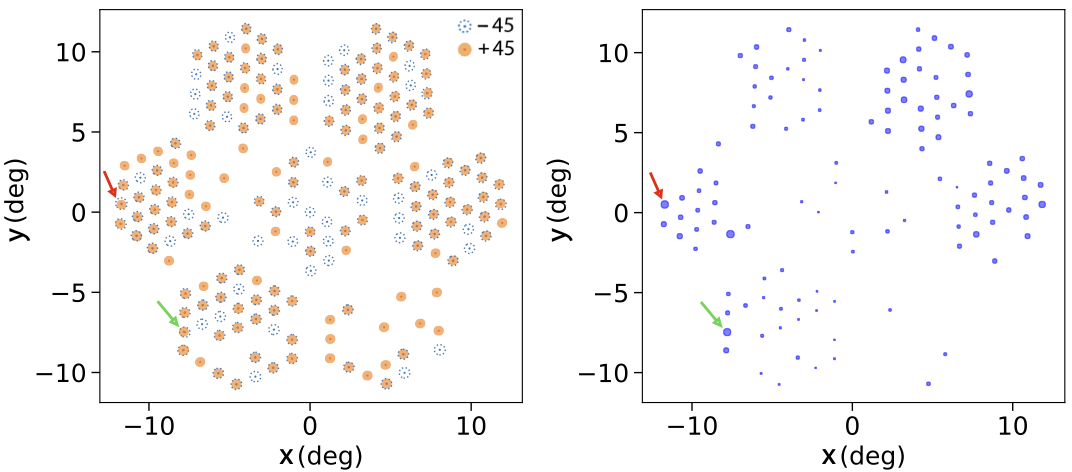}
\caption{{\textit{Left:}} Per-detector beam shapes and orientations as projected onto the sky represented by ellipses whose major and minor axes correspond to the FWHM beam widths. The $+45$\dg{} and $-45$\dg{} oriented detectors are shown in orange and blue, respectively. {\textit{Right:}} The blue filled circles are centered at the location of feedhorns that have fully operational pairs of detectors and represent the differential beam shapes within a co-located $\pm$45\dg{} detector pair, the area of the circle being proportional to the extent of beam mismatch as defined in Equation~\ref{equ:beam_mismatch}. The two pairs with the largest mismatch in beam shapes are indicated by the red and green arrows.}
\label{fig:beam_ellipticities}
\end{figure*}

In our optical design, off-axis detectors are expected to exhibit slightly asymmetric beams described by an elliptical shape with a Gaussian profile~\cite[e.g.,][]{1998A&AS..130..551B}. The left panel of Figure~\ref{fig:beam_widths} plots the ratio of the measured FWHM along the major and minor axes, $\sigma_\mathrm{maj}$/$\sigma_\mathrm{min}$ as a function of $\sigma_\mathrm{min}$. We define the beam eccentricity as
\begin{equation} 
    e = \sqrt{\frac{\sigma_\mathrm{maj}^{2}-\sigma_\mathrm{min}^{2}}{\sigma_\mathrm{maj}^{2}}}.
\end{equation}
While approximating the beam to be circular with a Gaussian profile can introduce systematic errors at angular scales
comparable to the beam FWHM~\citep{2002PhRvD..65f3003F,Burigana2001}, the scan strategy adopted by CLASS incorporating rotation about boresight makes the effective beam nearly circular (see Section~\ref{subsec:instrument_beam}). For each detector, we define an effective FWHM as
\begin{equation}
    \sigma_\mathrm{eff} = \sqrt{2\frac{\sigma_\mathrm{maj}^{2}\times\sigma_\mathrm{min}^{2}}{\sigma_\mathrm{maj}^{2}+\sigma_\mathrm{min}^{2}}}.
    \label{equ:effective_fwhm}
\end{equation}
The middle and right panels of Figure~\ref{fig:beam_widths} show histograms of $\sigma_\mathrm{eff}$ and the measured major axis orientation of the beams, respectively. The histograms are plotted per focal plane module.

Beam mismatch due to differential shape and gain between the beams of detectors with orthogonal polarization axes sharing a common feed can introduce systematics in polarization analysis~\citep{Fluxa20,2008PhRvD..77h3003S} if not properly accounted for. However, since our cosmological analysis pipeline works with individual detectors, the impact of these systematics is negligible. Nevertheless, we study these for the operational detector pairs since we use pair-differencing for probing certain instrumental systematics. Among the 319 detectors that detect Jupiter with S/N greater than 20, we have 115 operational pairs. Normalized per-detector gain calibration factors, equivalent to the end-to-end (i.e., detector plus telescope optics) optical efficiency of each detector, are obtained from the fitted peak amplitude of the Jupiter signal measured by each detector corrected for beam dilution. In Figure~\ref{fig:diff_gain} we plot the quantity $\epsilon_\mathrm{-}$ versus $\epsilon_\mathrm{+}$ for each operational detector pair, where $\epsilon_\mathrm{-}$($\epsilon_\mathrm{+}$) refers to this gain calibration factor of the $-45$\dg{}($+45$\dg{}) oriented detector. The large spread in the plots is a result of variations in detector efficiency~\citep[see][]{2022ApJ...926...33D} and emphasizes the importance of accurate calibration of the per-detector end-to-end efficiencies. New $90\,\mathrm{GHz}$ detectors 
 installed on the telescope in 2022 have demonstrated uniformly high efficiencies~\citep{10086549}.

The left panel of Figure~\ref{fig:beam_ellipticities} shows the beam shapes as defined by fitting a two-dimensional ellipse to the per-detector measured beam maps. The right panel shows the mismatch between measured beam shapes of detectors sharing a common feed. We quantify the fractional beam mismatch as
\begin{equation} 
     \delta B = \frac{2\int_0^{2\pi} |B_+(\theta) - B_-(\theta)| d\theta}{\int_0^{2\pi} [B_+(\theta) + B_-(\theta)] d\theta}
     \label{equ:beam_mismatch}
\end{equation}
\noindent where $B_+(\theta)$ and $B_-(\theta)$ are radial distances to ellipses fit to the $+45$\dg{} and $-45$\dg{} beams, respectively, in the $r$-$\theta$ plane at an angle $\theta$ referenced to the boresight axis of each feed. This metric uses the differences between shapes and orientations of the fitted elliptical beam envelopes as a proxy for beam mismatch. The average beam mismatch between co-located $\pm$45\dg{} detectors is 4.8\% with a scatter of 2.8\% across the focal plane. Of the 115 detector pairs, 6 have a mismatch $>10\%$. The various beam parameters including FWHM, eccentricity, beam orientation, enclosed solid angle within a 4\dg{} radius, and differential pointing and beam mismatch between the $\pm45$\dg{} oriented detectors are summarized per focal plane module in Table~\ref{tab:beam_params}. 

\begin{deluxetable*}{ccccccccc}
\tablenum{2}
\tabletypesize{\small}
\tablecaption{Summary of Era~2, $90\,\mathrm{GHz}$ beam and pointing parameters by focal plane module. Per module values are mean (standard deviation) of the measured per detector beam parameters. The aggregate values are the mean (standard deviation) over all measured detectors. The solid angles reported here are integrated measurements to a 4\dg{} radius. The aggregate beam FWHM and solid angle measurements are consistent with expectations from optical modeling of the telescope using the Zemax OpticStudio software. }
\label{tab:beam_params}
\tablehead{
\colhead{Module} & \colhead{N$_\mathrm{dets}$} & \colhead{FWHM} & \colhead{Eccentricity} & \colhead{Orientation} & \colhead{N$_\mathrm{det-pairs}$} &
\colhead{Differential pointing} & \colhead{Beam mismatch in pair} & \colhead{Solid Angle ($<$4\dg{})} 
}
\startdata
& & (deg) &  & (deg) & & (arcmin) & (\%) & ($\mu$sr)\\ 
\hline
1 & 52 & 0.636 (0.011) & 0.17 (0.03)  & $-81.0$ (17.3) & 22 & 0.8 (0.9) & 3.0 (2.9) & 142 (8) \\ 
2 & 20 & 0.645 (0.007) & 0.20 (0.04)  & $-74.4$ (14.4) & 5 & 0.7 (0.2) & 4.4 (2.1) & 152 (6) \\
3 & 47 & 0.629 (0.017) & 0.22 (0.04)  & $+45.2$ (10.1) & 19 & 1.5 (0.4) & 6.5 (3.1) & 146 (10) \\ 
4 & 34 & 0.607 (0.008) & 0.15 (0.05)  & $-16.9$ (33.8) & 9 & 0.7 (0.6) & 2.5 (1.2) & 127 (6) \\ 
5 & 53 & 0.623 (0.008) & 0.26 (0.04)  & $-46.2$ (6.4) & 21 & 1.7 (0.6) & 5.5 (2.0) & 134 (10)\\ 
6 & 52 & 0.611 (0.005) & 0.29 (0.03) & $+9.6$ (8.2) & 16 & 0.9 (0.4) & 3.3 (1.3) & 132 (13) \\ 
7 & 61 & 0.614 (0.010) & 0.31 (0.03) & $-18.2$ (7.0) & 23 & 1.6 (0.5) & 5.8 (1.6) & 133 (7) \\ 
\hline   
Aggregate & 319 & 0.622 (0.015) & 0.24 (0.07) & & 115 & 1.2 (0.7) & 4.8 (2.8) & 137 (12)\\ 
\enddata
\end{deluxetable*}

\vskip 5.8mm plus 1mm minus 1mm
\subsection{Effective Instrument Beam}
\label{subsec:instrument_beam}

\begin{figure}
\centering
\includegraphics[width=1\linewidth]{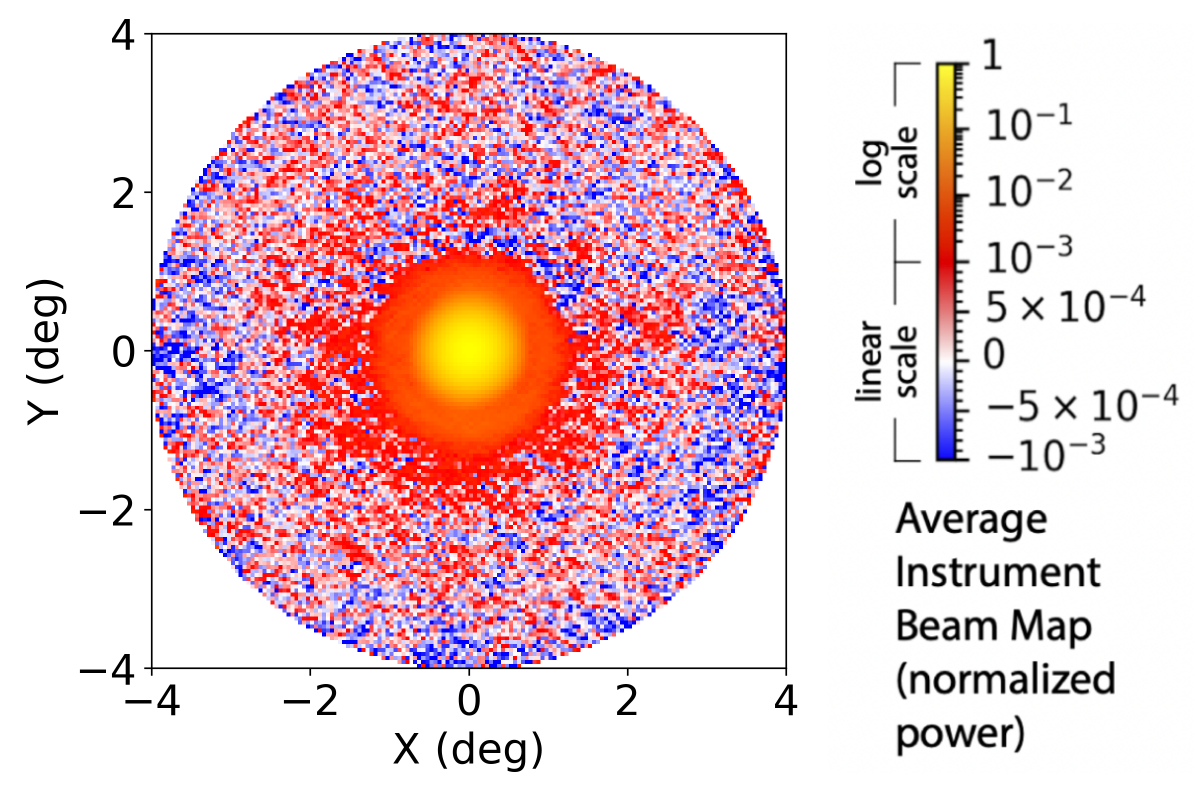}
\caption{\textbf{Instrument Beam}. Peak-normalized beam map to 4\dg{} in radius obtained by stacking beam measurements of 206 detectors from multiple dedicated scans of Jupiter. The color bar scale is logarithmic from 1 to $10^{-3}$, and linear from $10^{-3}$ to $-10^{-3}$ in order to display negative values. The map has a resolution of 0.05\dg{}.}
\label{fig:averaged_beam_map}
\end{figure}

For each detector, the set of scans accepted according to the criteria described in Section~\ref{sec:beam_meas}, are subjected to random sampling with replacement and averaged as described in Section~\ref{sec:beam_meas} to generate 100 bootstrapped samples. A set of 100 instrument beam maps are computed  by scaling these by the detector relative efficiencies and co-adding with inverse variance weighting. The effective instrument beam map is then the equal weighted average of the 100 sample instrument beam maps. When co-adding, we excluded detectors with (1) low end-to-end relative efficiency, (2) highly eccentric beams, or (3) an out-of-range voltage bias (when the TES is biased above $0.8R_\mathrm{N}$ or below $0.2R_\mathrm{N}$, where $R_\mathrm{N}$ is the TES normal resistance) for the majority of Jupiter scans. We define relative efficiency as $\epsilon_\mathrm{rel,i}$ = $\epsilon_\mathrm{i}$/$\epsilon_\mathrm{mean}$, where $\epsilon_\mathrm{i}$ is the absolute efficiency of the i\textsuperscript{$th$} detector and $\epsilon_\mathrm{mean}=0.4$ is the mean end-to-end efficiency over all detectors. Of the 319 optical detectors for which we have beam measurements as described in Section~\ref{sec:beam_meas}, we retained 270 detectors with $\epsilon_\mathrm{rel,i}>0.5$. We excluded 31 detectors with beam eccentricity $e>0.5$. Combined with the selection on $\epsilon_\mathrm{rel,i}$, this brings the number of acceptable detectors to 251. Finally, we excluded an additional 45 detectors with out-of-range voltage bias. 

\begin{figure}
    \centering
    \includegraphics[width=1\linewidth]{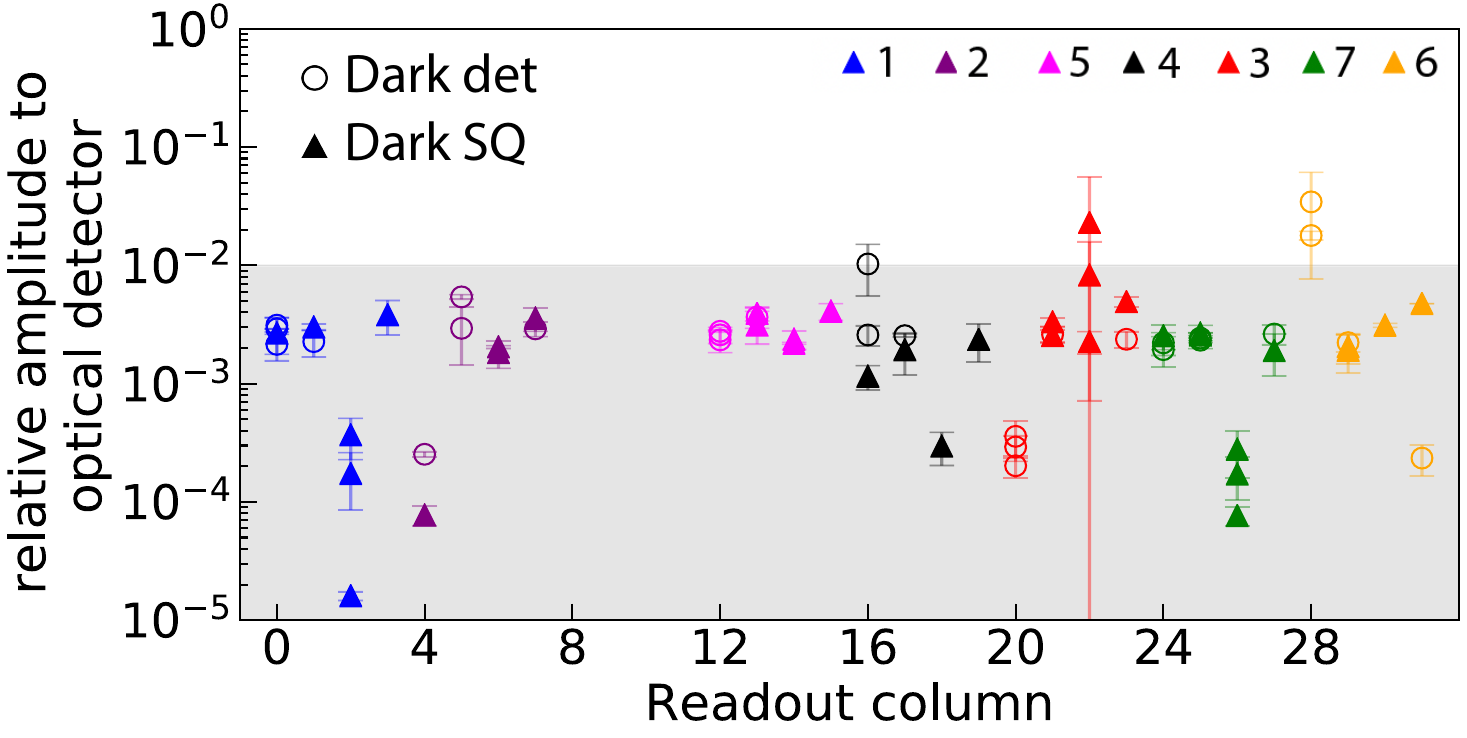}
    \caption{Electrical crosstalk per readout column inferred from dark detector and dark SQUID channels and their 1$\sigma$ uncertainties. Each color represents a module as indicated in the legend, and each module is read out by four columns. The grey region represents electrical crosstalk $<1\%$.}
    \label{fig:elec_crosstalk}
\end{figure}

The above selection criteria resulted in 206 detectors being used to compute the effective instrument beam map shown in Figure~\ref{fig:averaged_beam_map}. Due to the left-right symmetry of the optics, averaging per-detector maps from across the focal plane circularizes the instrument beam map. When forming the effective cosmology beam from the instrument beam, boresight rotation and observations of the same point rising and setting (i.e., scan cross-linking) further ensure that the effective beam for CMB observations is nearly circular. Data from the seven boresight orientations contribute with nearly equal weights to the survey map. Therefore, we approximate our effective beam by an azimuthally averaged radial beam profile. The effective instrument beam map is used to generate the 1D beam profile and its uncertainties. The measured FWHM of this profile is  $0.620\pm0.003$\dg{}.

To estimate the systematic impact of the baseline removals during beam processing described in Section~\ref{sec:beam_meas} on the beam profile estimation, we generated simulated timestreams for the 206 selected detectors based on the timestamps of the Jupiter scans used for beam measurements. A modeled beam described later in Section~\ref{subsec:cosm_beam} was used for the simulations. The beam analysis was repeated on these simulated timestreams both with and without the baseline removal steps. The difference between the resulting averaged 1D beam profiles was used to debias the measured effective instrument 1D beam. If not corrected, this systematic would bias the beam window function (discussed later in Section~\ref{subsec:beam_profile_and_window_function}) higher by $<2\%$ for $\ell\lesssim100$, and $\sim2\%$ for $\ell>100$.  The impact of debiasing on the measured FWHM is negligible.

\vskip 5.8mm plus 1mm minus 1mm
\subsection{Electrical Crosstalk}
\label{sec:crosstalk}
Crosstalk refers to spurious coupling of any neighboring detector’s signal to the signal measured by a particular detector and can be induced by imperfect isolation in the readout architecture resulting in current in one TES loop coupling to the current in another TES loop.

\begin{figure*}
    \centering
    \includegraphics[width=1\linewidth]{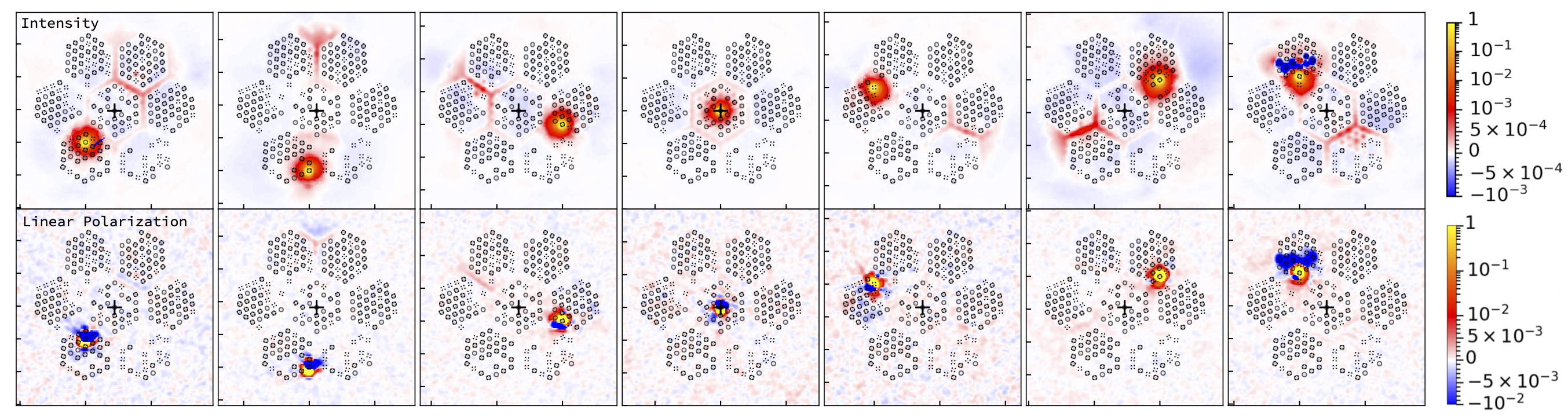}
    \caption{\textbf{Examples of ghosting beam}. Selected detector-centered peak-normalized Moon maps in intensity (top) and linear polarization (bottom) with their locations on the focal plane indicated by the array plot in the background. These maps span 30\dg{} $\times$ 30\dg{} about the center of the focal plane indicated by the black `+' symbol. Sub-percent level reflections (relative to the peak) off the opposite side of the focal plane are seen in these maps. For instance, in the top right panel, the bright circular spot with unit amplitude is the main beam. On the other side of the focal plane, ghosts trace the field stop between detector modules. The negative blue dots near the main beam come from electrical crosstalk. }
    \label{fig:ghosting}
\end{figure*}

The readout scheme for CLASS is implemented by eight time-division multiplexing chips per detector module. Each chip contains eleven SQUIDs, and two such chips are paired in a ``column'' to multiplex 22 SQUIDs. The SQUIDs are sequentially activated by flux switches to read out one detector at a time per readout column. Each module is read out by four columns across which 22 ``rows'' are read out in sequence~\citep{daha18,reintsema2003}. The electrical crosstalk, as inferred by measuring the signal in dark detector time-streams when an optical detector in the same column scans across the Moon, is $<1\%$ for the majority of the 28 readout columns. Figure~\ref{fig:elec_crosstalk} shows the limits for each readout column. For two of the readout columns, this crosstalk was estimated at $\sim 2\%$. The affected detectors are in modules 3 and 6. Detectors being read out through these columns were excluded when computing the averaged instrument beam. However, this criterion does not exclude any additional detectors from the list of 206 detectors being considered for the averaged beam. 

\subsection{Sidelobes caused by internal reflections}
\label{subsec:far_sidelobe}
Minimizing pick-up through potential far-sidelobes of the telescope beam is critical for control over systematics. Unmitigated, this could lead to contamination from the ground and bright sources on the sky far away from the telescope's field-of-view. CLASS uses a co-moving baffle that rotates with the telescope boresight and follows the scanning (see Figure~\ref{fig:bs_rotation}). It comprises the telescope structure housing the fore-optics and receivers, which we call the ``cage'' and an ``extension'' on the sky side of the circular telescope aperture. The extension is designed to intercept light from the horizon up to an elevation angle of $\sim20$\dg{} as seen from the top of the VPM mirror during CMB survey scans performed at an elevation of 45\dg{} and 0\dg{} rotation about boresight. Therefore, the line-of-sight from the highest ground feature (mountain peak at $\sim14$\dg{} elevation at the edge of the under-illuminated VPM mirror) is intercepted by it. The extension has rolled top edges ($76.2\,\mathrm{mm}$ in diameter) to reduce diffraction around sharp edges. 

The interior of the telescope cage is covered with a broadband microwave absorber (i.e., Eccosorb HR10 carbon-loaded open-cell polyurethane foam) to terminate sidelobes and stray light. Stray light or ground signals not intercepted by the extension and/or making it to the VPM mirror through indirect paths will either terminate on the absorber covered interiors of the cage or be reflected back out of the extension (see Figure~\ref{fig:ray_schematic}). Experimenting with various absorber configurations, we found that covering the interior surfaces of the cage with absorber significantly
reduces ground pick-up in the direction of the mountain (i.e., stratovolcano Cerro Toco), while the addition of absorptive material on the inside of the extension has negligible impact on the ground pick-up. From the baffle and optical configuration this behavior is consistent with the anticipated instrument response. In addition, absorber-coated
corrugations along the inside surface of the cryogenic receiver's optics tube are employed to dampen grazing incidence reflections.

\begin{figure*}
\centering
\includegraphics[width=1\linewidth]{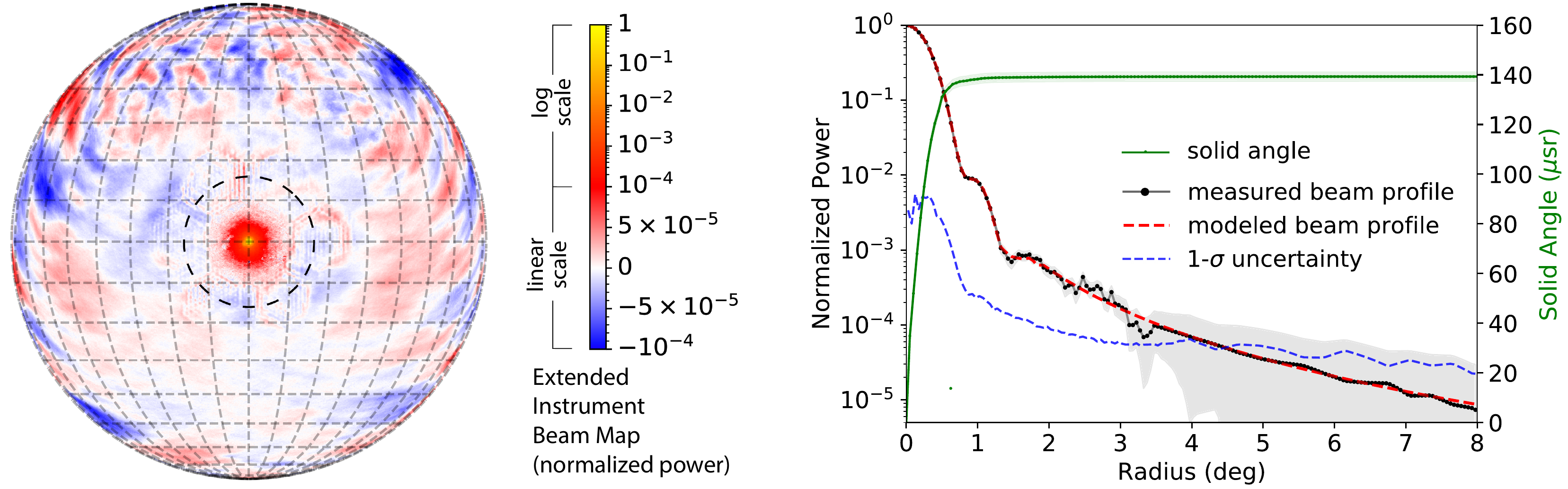}
\caption{\textbf{Extended intensity beam map}: {\textit{Left:}} Peak-normalized instrument beam, obtained from a combination of the average beam map from dedicated Jupiter scans and the detector-centered Moon map generated from survey data. The color bar scale is logarithmic from 1 to $10^{-4}$, and linear from $10^{-4}$ to $-10^{-4}$ in order to display negative values. The negative values in this intensity beam map are a result of mapping artifacts and measurement noise.} The 8\dg{} radius dashed circle indicates the extent of the measured 1D beam profile plotted in the right panel. {\textit{Right:}} Azimuthally averaged beam profile to 8\dg{}. The shaded error region represents 1$\sigma$ uncertainties, also indicated by the blue line. The red line is the best-fit model (Equation~\ref{equ:beam_model}). The green line shows the solid angle enclosed as a function of angular distance. The larger scatter in the measured beam profile between 1.5\dg{} and 4\dg{} is a result of different bin sizes used between the planet beam map and the survey Moon map.
\label{fig:extended_averaged_beam_map}
\end{figure*}

While far-sidelobes can be modeled using physical optics simulations~\citep[e.g.,][]{2018SPIE10708E..2LG}, for this work we leverage survey Moon scans to characterize far-sidelobes. Per-detector maps of the Moon in detector-fixed coordinates are generated using an adapted version of the map-making algorithm~\citep{Li23} used for CMB maps. Maps in intensity and linear polarization for a selected number of detectors are shown in Figure~\ref{fig:ghosting} with an array plot overlaid to indicate the detector's location on the focal plane. Sub-percent level reflections relative to the peak amplitude off the opposite side of the focal plane are seen in the intensity maps. These reflections also appear in polarization maps with relative amplitudes similar to that in intensity. Some readout channels are more susceptible to electrical crosstalk as discussed in Section~\ref{sec:crosstalk}. For example, in the rightmost panel the positive and negative spots coincident with neighboring detector locations are a result of electrical crosstalk. For a given detector, the region from which the reflection peaks is diametrically opposite to itself about the center of the focal plane. We model these reflections using a functional form obtained by fitting to the data from about 200 individual detector maps. These per-detector models are not accurate enough to be included in the instrument beam model; instead we use them in simulated maps to assess their impact on the inferred CMB power spectra, see Section~\ref{sec:systematics}. We suspect that these reflections are caused by a small fraction of light reflecting off the field stop, which is an absorber-coated copper structure behind and in between the feedhorns (see Figure~\ref{fig:ray_schematic}) and again reflecting off some optical element (e.g., AR coated lenses, filters, vacuum window) near the cold stop back towards the focal plane. Efforts to further understand and narrow down the source of this ghosting through geometric optics simulations are currently underway. Additionally, some higher efficiency detectors appear to see reflections off other feedhorns located in the region of the focal plane diametrically opposite to itself about the focal plane center when those feedhorns are pointing at the Moon. This effect is more prominent when the detectors coupled to those feedhorns have lower efficiency. 

\vskip 5.8mm plus 1mm minus 1mm
\subsection{Extended Beam} 
\label{subsec:cosm_beam}

The averaged beam map from the dedicated Jupiter scans is combined with a Moon-centered survey map generated using data from all 206 detectors to produce an extended beam map out to large angles. The Moon was deconvolved from the survey map; the central region out to 3\dg{} was masked; and the resulting map was then stitched with the planet beam map. The planet beam map is used for the central region because the Moon's signal nearly saturates many of the detectors. At large angles, away from the peak of the Moon signal, contamination from 1/$f$ low-frequency fluctuations is non-negligible. To mitigate this, the overlapping annulus with the planet beam map 3\dg{}$< \theta <$4\dg{} was used to normalize the survey map amplitude to a common reference following a similar prescription as in \cite{2020ApJ...891..134X}. The left panel of Figure~\ref{fig:extended_averaged_beam_map} shows the extended instrument intensity beam, where features from the ghosting discussed previously can be seen. Beyond these ghosting features with amplitudes on the order of 10$^{-3}$ seen within a radius of about 12\dg{} around the center, an upper limit on the amplitude of far-sidelobes on the order of 10$^{-4}$ is inferred from the extended instrument intensity beam map. With approximately equal amount of data at each boresight, the ghosting features are essentially spread out in azimuth and hence their impact is reduced in the effective cosmology beam. The map data are binned to generate an azimuthally averaged radial beam profile shown in the right panel of Figure~\ref{fig:extended_averaged_beam_map}. This profile extends to 8\dg{}.  At angles larger than 8\dg{}, where the beam level falls below $10^{-5}$ of the peak, residual background fluctuations confuse the asymptotic part of the beam profile. 

For modeling the beam, Zernike polynomials provide a natural basis in Fourier space. In real space, under the assumption of azimuthal symmetry, we model the radial beam profile with basis functions that are proportional to inverse Fourier transforms of the Zernike polynomials~\cite[e.g.,][]{hass13}. The illumination of the optics is controlled by a cold circular aperture stop, which is at an image of the VPM created by the primary and secondary mirrors, leading to an Airy pattern intensity beam shape for monochromatic radiation. With some approximations, it can be shown that the envelope of this Airy pattern asymptotically decays as $1/\mathrm{sin}^{3}\theta$ at large angles, where $\theta$ is the angle away from the direction of the peak. Therefore, we fit a model to the radial profile $B(\theta)$ defined as~\cite[e.g.,][]{hincks2010atacama,hass13,2022JCAP...05..044L}:
\begin{equation} 
    B(\theta)= 
\begin{dcases}
    \sum_{n=0}^{n_\mathrm{mode}-1}\frac{a_{n}J_{2n+1}(\theta \ell_\mathrm{max})}{\theta \ell_\mathrm{max}},& \text{for } \theta\leq \theta_{0}\\
    \frac{b}{\mathrm{sin}^{3}\theta}+S(\theta), & \text{for } \theta > \theta_{0}
\end{dcases}
\label{equ:beam_model}
\end{equation} 
\noindent where $J_{2n+1}$ refers to the Bessel function of the first kind of order $2n+1$, $\theta_{0}$ is the radius at which the model is switched, and $S(\theta)$ is an additional scattering term given by~\cite{Ruze:1996} and can be approximated for our purposes as
\begin{equation} 
    S(\theta) \approx \bigg(\frac{2\pi c}{\lambda}\bigg)^{2} \bigg[\frac{2\pi(1+\cos\delta) \epsilon}{\lambda}\bigg]^{2} e^{-(\frac{c\pi\sin\theta}{\lambda})^{2}}
\label{equ:ruze}
\end{equation} 

\noindent where $\epsilon$ is an RMS surface roughness and $c$ is a correlation length that characterizes random errors on a reflector surface, $\delta$ is the offset angle of the reflector. Mirror surface aberrations scatter the field from the forward direction into the side lobe. The reduction in the peak gain due to this diffuse scattering is, however, expected to be negligible in our case. For modeling the central part of the beam profile, the parameters $\ell_\mathrm{max}$, $\theta_{0}$, and $n_\mathrm{mode}$ were chosen such that the reduced $\chi^{2}$ of the fit approached unity with the minimum possible number of modes, resulting in $n_\mathrm{mode}=10$, $\theta_{0}=1.875$, and $\ell_\mathrm{max}=749$ with a reduced $\chi^{2}$ of 1.2. While running the fits, the parameter $n_\mathrm{mode}$ was constrained to be less than a third of the number of data points being fitted in the central part of the beam (i.e., main-beam); $\theta_{0}$ was constrained to be greater than the radius at which the measured beam profile first drops $20\,\mathrm{dB}$ below its peak. The right panel of Figure~\ref{fig:extended_averaged_beam_map} shows a model fit to the measured radial profile. The profile was fit to 8\dg{} with a reduced $\chi^{2}$ of 1.7 for the combined fit including the central region and the asymptotically decaying region. For a beam profile $B(\theta)$, the corresponding solid angle $\Omega$ is given by
\begin{equation}
\label{equ:beam_sa}
    \Omega = 2\pi \int_{0}^{\pi} B(\theta)\sin \theta d\theta. \
\end{equation}
The solid angle enclosed as a function of angular distance $\theta$ is also plotted in the right panel of Figure~\ref{fig:extended_averaged_beam_map}. The enclosed solid angle is $138.7\pm0.6$(stats.)$\pm1.1$(sys.)$\mu{\rm sr}$ within a radius of 4\dg{}. The statistical uncertainty is obtained from the sample variance of the solid angles computed from the 100 bootstrapped beam profiles~(Section~\ref{subsec:instrument_beam}). The systematic uncertainty is the uncertainty associated with the debiasing of the 1D beam profile. The additional solid angle between 4\dg{} and 8\dg{} is $<$0.1\%. 

\vskip 5.8mm plus 1mm minus 1mm
\subsection{Beam Window Function}
\label{subsec:beam_profile_and_window_function}
\begin{figure}
    \centering
    \includegraphics[width=1.0\linewidth]{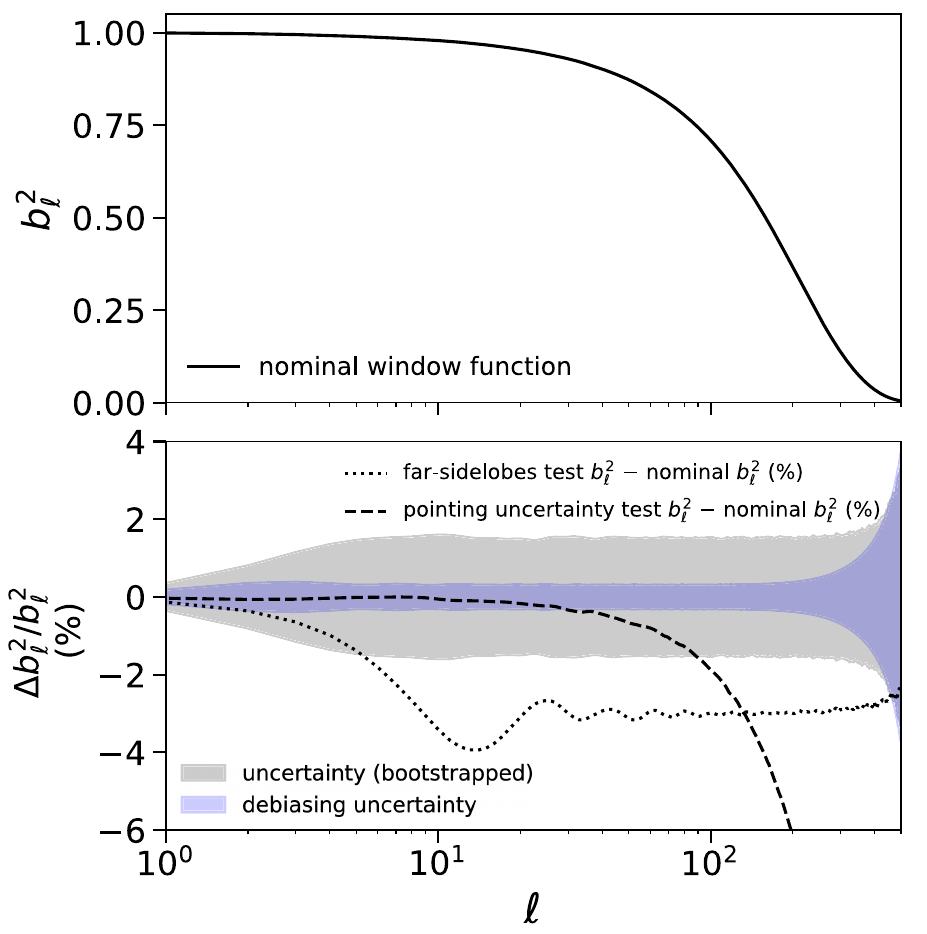}
    \caption{\textbf{Beam window function.} {\textit{Top:}} The nominal window function $b_\mathrm{\ell}^\mathrm{2}$ unit normalized at $\ell=0$ plotted as a function of multipole $\ell$. {\textit{Bottom:}} The grey band shows the relative uncertainties $\Delta b_\ell^2/b_\ell^2$, estimated from simulated window functions computed from bootstrapped beam profiles. Uncertainties from the 1D beam debiasing are also plotted. Residuals (in \%) between test window functions computed to understand impact of pointing uncertainties and far-sidelobes and the nominal window function are also plotted. }
    \label{fig:beam_window_func}
\end{figure}

The harmonic transform of the beam in $\ell$ space, $b_\ell$ and the associated beam window function $b_\ell^2$ were computed from the measured and debiased radial beam profile using the relation
\begin{equation}
\label{equ:beam_transform}
    b_{\ell} = \int d\Omega\; B(\theta)P_{\ell}(\cos \theta),
\end{equation}
\noindent where $P_\ell$ is the $\ell^\mathrm{th}$ Legendre polynomial. For computing the integral to $\theta = 90$\dg{}, the model fit to the beam profile in Section~\ref{subsec:cosm_beam} was extrapolated to 90\dg{}. Figure~\ref{fig:beam_window_func} shows the window function $b_\ell^2$ unit normalized at $\ell=0$ and its uncertainties. To estimate the uncertainties in the window function, the 100 bootstrapped beam profiles (Section~\ref{subsec:instrument_beam}) were used to compute 100 simulated window functions. The uncertainties on the window function were then estimated as the sample variance of the simulated window functions. The unit-normalized window function $b_\mathrm{\ell}^\mathrm{2}$ drops to 0.93 at $\ell=30$, 0.82 at $\ell=70$, 0.71 at $\ell=100$, and 0.14 at $\ell=300$. The relative uncertainty $\Delta b_\ell^2/b_\ell^2$ is plotted in the lower panel of Figure~\ref{fig:beam_window_func}. 

Drifts in pointing per period, i.e., between updates in the pointing model, have the effect of broadening the beam. A broadened beam profile informed by the measured pointing deviations in Section~\ref{subsec:pointing_data} and assumption of Gaussian-distributed pointing errors with residual at the level of 3$^{\prime}$ was used to compute a test window function. The resulting bias in the beam window function $b_\mathrm{\ell}^\mathrm{2}$ due to the broadened beam profile is plotted in the lower panel of Figure~\ref{fig:beam_window_func} and is $<2\%$ for $\ell\lesssim100$, and $\sim12\%$ at $\ell\sim300$. This effect is being further investigated using maps generated from the survey data. Also plotted here are uncertainties associated with the debiasing of the 1D beam profile inferred from the simulations described in Section~\ref{subsec:instrument_beam}. Additionally, to estimate how inaccuracies in modeling of the beam at large angles and far-sidelobes might affect the beam window function, we computed a test window function from a hypothetical profile where the level of the peak-normalized beam is fixed at $10^{-5 }$ for 8\dg{}$< \theta <$20\dg{}. The residuals between this test $b_\ell^2$ and the nominal $b_\ell^2$ are also plotted for this unlikely worst case scenario. We are considering physical optics modeling of the telescope for accurately constraining far-sidelobes for future work.

\vskip 5.8mm plus 1mm minus 1mm
\section{Polarization Angle}
\vskip1sp
\label{sec:pol_ang}

In general, knowledge of the absolute detector polarization angles on the sky is important for controlling systematic effects. In particular, $E\rightarrow B$ leakage due to polarization angle errors would directly hamper the ability to measure the primordial B-modes. This leakage is quantified by EB correlation, which is not expected in the standard model of cosmology. Therefore, one approach to mitigate systematics due to miscalibrated polarization angles uses a self-calibration technique by nulling the $C_\mathrm{\ell}^\mathrm{EB}$ spectra \citep{2013ApJ...762L..23K, 2020PhRvD.102h3504B}. 

\begin{deluxetable}{@{\extracolsep{2pt}}cccccc}
\tablenum{3}
\tablecaption{CLASS measurement of Tau~A polarization angle $\psi$ based on preliminary $90\,\mathrm{GHz}$ maps, compared with \textit{Planck} HFI, WMAP, and IRAM measurements. The IRAM numbers are from maps made with 27~arcsec resolution and convolved with a 10~arcmin simulated beam. }
\label{tb:TauA_angle}
\tablehead{
\colhead{} & \colhead{Freq} &
\colhead{beam size} & \colhead{$\psi$} & \multicolumn{2}{c}{Uncertainty (deg)} \\
\cline{5-6}
\colhead{} & \colhead{(GHz)} & \colhead{(arcmin)} & \colhead{(deg)} & \colhead{statistical} & \colhead{systematic} 
}
\startdata
WMAP & 93 & 13 & 148.9 & 0.7 & 1.5  \\
\textit{Planck} HFI & 100 & 10 & 150.1 & 0.16 & $<$1.0 \\
IRAM & 90 & 10 & 148.8 & 0.2 &  \\
CLASS$^{*}$ & 90 & 37 & 149.6 & 0.2 & \\
\enddata 
\tablecomments{$^{*}$This work}
\end{deluxetable}

\begin{figure*}
    \centering
    \includegraphics[width=0.85\linewidth]{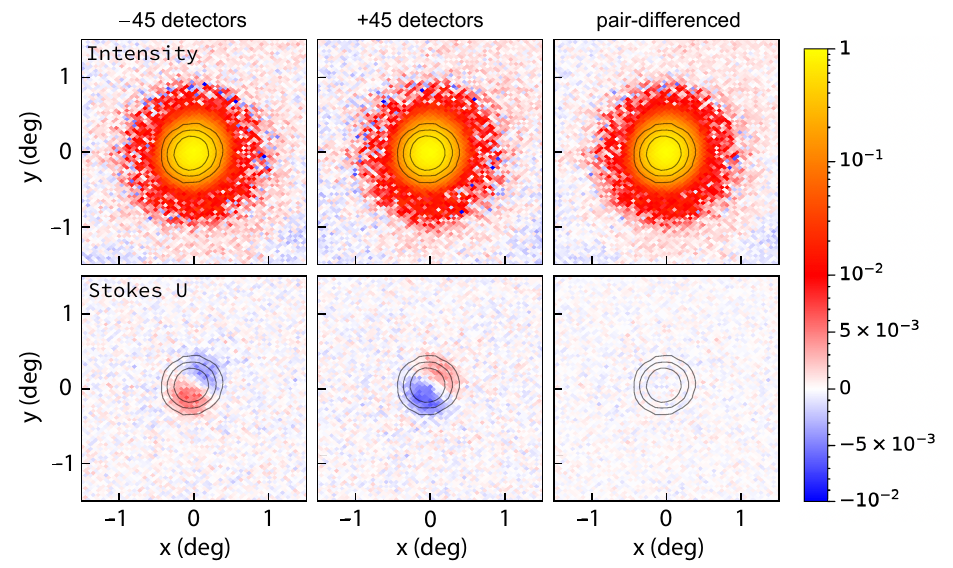}
    \caption{\textbf{$T\rightarrow P$ leakage probed using dedicated Jupiter scan data}. {\textit{Top row}}: Peak-normalized average intensity maps spanning 3\dg{}$\times$3\dg{} generated from single-detector ($-45$\dg{} and $+45$\dg{} oriented detectors averaged separately) and pair-differenced raw (i.e., not demodulated) Jupiter scan data. The black contours are plotted at 3, 6, and $10\,\mathrm{dB}$ below the peak. {\textit{Bottom row}}: Corresponding average linear polarization maps (Stokes $U$ in the instrument frame) generated from single-detector and pair-differenced demodulated Jupiter scan data. These maps are normalized by the peak value of the intensity map. The contours from the corresponding intensity maps are overplotted.   Jupiter is not expected to have a polarized signal at $90\,\mathrm{GHz}$. However, a dipolar $T\rightarrow P$ leakage component is seen in the Stokes $U$ maps made from averaging over only $-45$\dg{} or $+45$\dg{} oriented single detectors. In addition, there is also a monopole leakage component in the average single-detector Stokes $U$ maps. No leakage signal is seen in the Stokes $U$ map made from pair-differenced detector data in the bottom right panel. }
    \label{fig:ItoP_demodpd}
\end{figure*}

However, cosmic polarization rotation could theoretically be sourced by, for example, primordial magnetic fields \citep{2016A&A...594A..19P} or parity-violating extensions of the standard model \citep{2010PhRvD..81l3529G, 2016A&A...596A.110P}. The resulting cosmic birefringence would lead to a non-vanishing EB correlation as would polarized Galactic foregrounds \citep{2016A&A...586A.133P}, which would be degenerate with a miscalibration of detector polarization angles. Hence, CMB experiments are developing alternative methods, e.g., using the Crab Nebula (Tau~A) \citep{2022ApJS..258...42S,2020EPJWC.22800003A} or polarized Galactic foregrounds \citep{2019PTEP.2019h3E02M,  2020PTEP.2020j3E02M}, artificial polarization calibration sources \citep{2022SPIE12190E..1XC, dunner20drone, 2017JAI.....640008N, bicep15}, or map based deprojection \citep{2020JLTP..199..824S} to calibrate the detector polarization angles. Polarization rotations induced by the optics can also be modeled using polarization sensitive ray tracing software~\citep{2016SPIE.9914E..2TK}.

For CLASS, the relevant angle is related to the orientation of the detector antenna probes with respect to the VPM and the projected angle of the VPM wires onto the sky. These are initially estimated from our knowledge of the telescope geometry and serve as input to the VPM modulation functions used to demodulate the raw data prior to generating maps of the sky in Stokes $I$, $Q$, and $U$. We then use measurements of the polarized signal from the Crab nebula (Tau~A) in these survey maps for a consistency check of the instrument polarization angle on the sky to a precision of a few degrees. The emission from the Crab nebula has been shown to be synchrotron dominated at millimeter wavelengths with a simple power-law frequency spectrum~\citep{2010ApJ...711..417M, 2011ApJS..192...19W}.

Since Tau~A only rises to $\sim$45\dg{} elevation at the CLASS site, only a subset of detectors in the lower half of the focal plane measure Tau~A with the requisite boresight angle coverage during the CMB survey scans. The polarization angle $\psi$ is defined as
\begin{equation} 
    \psi = \frac{1}{2}\mathrm{tan}^{-1}\bigg(\frac{U}{Q}\bigg). 
\end{equation}
Following the IAU convention, the Tau~A polarization angle from a preliminary $90\,\mathrm{GHz}$ survey map is measured to be $149.6\pm0.2$\dg{} in equatorial coordinates. These measurements are compared with previous measurements from WMAP~\citep{2011ApJS..192...19W} at $93\,\mathrm{GHz}$, \textit{Planck} HFI~\citep{2020EPJWC.22800003A} at $100\,\mathrm{GHz}$, and the IRAM 30~m telescope~\citep{2010A&A...514A..70A} at $90\,\mathrm{GHz}$ as listed in Table~\ref{tb:TauA_angle}. When considering the uncertainties of each experiment, the CLASS $90\,\mathrm{GHz}$ measurement is consistent with the WMAP and \textit{Planck} measurements. 

\begin{figure}
    \centering
    \includegraphics[width=1.0\linewidth]{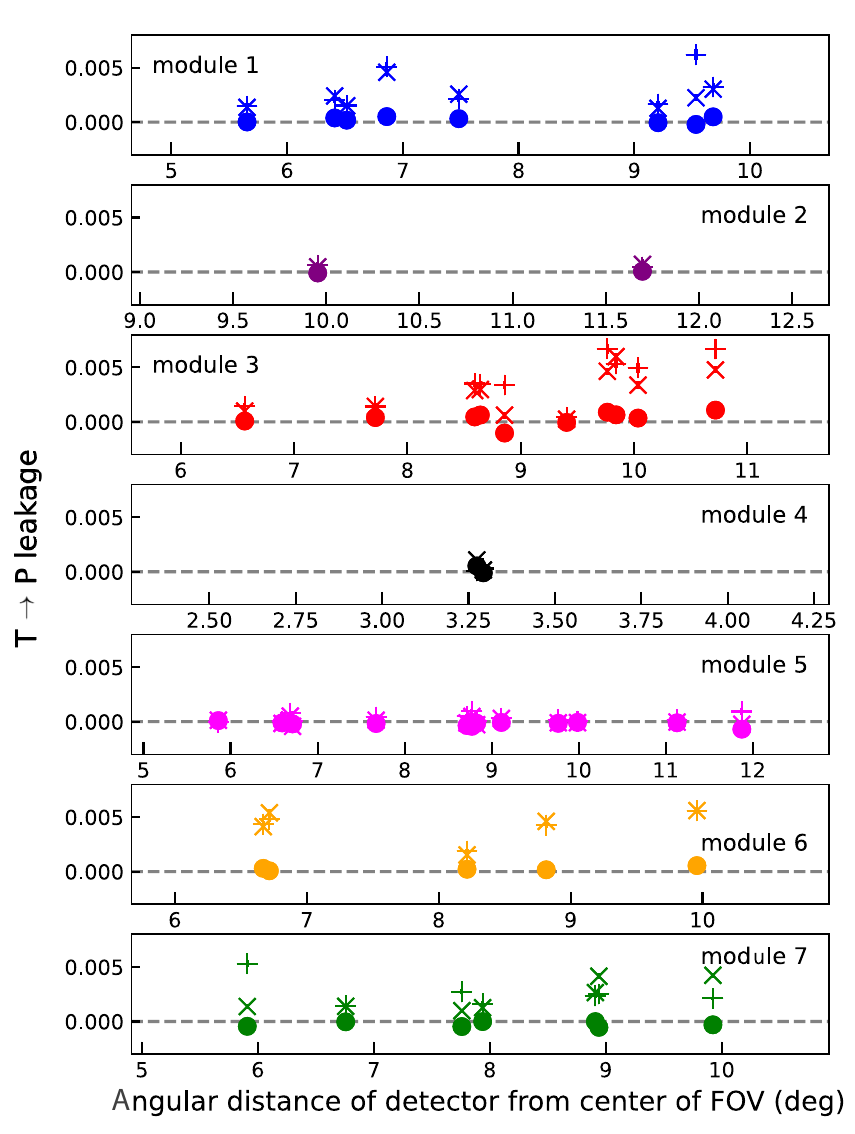}
    \caption{\textbf{$T\rightarrow P$ leakage probed using survey Moon scan data}. Upper limits of 95\% confidence interval on $T\rightarrow P$ leakage estimated from per-detector survey Moon maps plotted with the detector angular distance from the FOV center along the $x$-axis, and grouped by module. These are estimates for 49 pairs of detectors that are not saturated by the Moon's signal and include contribution from the Moon's intrinsic polarization. For module~4, data from only two detector pairs were usable for this analysis. The `$\times$,' `+,' and \markerzero\,\ represent measurements from $-45$\dg{}, $+45$\dg{}, and pair-differenced detectors, respectively. }
    \label{fig:ItoP_Moon}
\end{figure}

\vskip 5.8mm plus 1mm minus 1mm
\vskip1sp
\section{Temperature-to-Polarization Leakage}
\vskip1sp
\label{sec:TtoP_leakage}
The demodulated detector response to a linearly polarized sky signal, modulated by the VPM, can be expressed in the detector $Q$/$U$ coordinate system as given by~\cite{harr18}:
\begin{equation}
    U = U_\mathrm{in}+\lambda_U I_\mathrm{in} + \textnormal{VPM grid emission terms}
    \label{eq:real_p_demod}
\end{equation}

\noindent where the detector $U$ axis is defined along the orientation of the $+45$ detector; the $-45$ detector then measures $-U$. Here, $U_\mathrm{in}$ is the intrinsic polarization while $\lambda_U$ is the $T\rightarrow P$ leakage term. In this work, we use data from dedicated observations of Jupiter that are undertaken for beam characterization and Moon survey data to obtain an estimate of the $T\rightarrow P$ leakage.

\subsection{Jupiter observations}
The data from dedicated Jupiter scans are useful for probing the instrumental $T\rightarrow P$ leakage since no polarization signal is expected from Jupiter. After calibrating, filtering, and downsampling the raw TODs, we demodulate the data following an established demodulation technique~\citep{harr21,Li23}. Figure~\ref{fig:ItoP_demodpd} shows Stokes $I$ maps along with maps of Stokes~$U$ in the instrument frame made from $-45$ detectors, $+45$ detectors (i.e., single-detector), and their difference (i.e., pair-differenced). These are average maps obtained by stacking multiple maps from individual Jupiter scans over well-behaved detector pairs. No signal was detected above the noise level in the pair-differenced Stokes $U$ map. When compared to the peak amplitude in the Stokes $I$ map, this implies a 95\% confidence upper limit of 5.5$\times$10$^{-4}$ on the $T\rightarrow P$ leakage fraction. Single-detector Stokes $U$ maps indicate a dominant dipolar $T\rightarrow P$ leakage component on the order of $4.3\times$10$^{-3}$ in addition to a monopole leakage on the order of $1.7\times$10$^{-3}$. The dipolar leakage results from a varying tilt of the flat movable mirror of the VPM relative to the fixed wire grid as a function of distance between the wire grid and the mirror. This has been verified through modeling. This has important implications in terms of beam systematics in survey maps containing data from detectors that do not have an operational pair. The impact of these systematics on CMB power spectra estimation are studied using simulations and presented in Section~\ref{sec:systematics}. Quadrupolar leakage could be the most detrimental of the various systematics considered for $B$-mode power spectrum estimation~\citep{2008PhRvD..77h3003S}. We find no evidence of a quadrupolar $T\rightarrow P$ leakage component in these maps.

\vskip 5.8mm plus 1mm minus 1mm

\subsection{Moon observations}
The $T\rightarrow P$ leakage is also probed using per-detector survey Moon maps in intensity and polarization. Since many detectors are either saturated or driven to a non-linear regime by the strong Moon signal, we exclude these detectors from this analysis based on their response to the Moon signal. The measured Moon polarization maps contain a monopole component along with dipole and quadrupole components similar to what was seen at $40\,\mathrm{GHz}$ by~\cite{2020ApJ...891..134X}. While the monopole component is likely to be a combination of the Moon’s intrinsic polarization and the $T\rightarrow P$ leakage, the quadrupole pattern is expected due to refraction from the Moon's regolith. 

Average Moon polarization maps for each single-detector and pair-differenced detector were projected to Gauss–Hermite functions similar to the treatment by~\cite{2020ApJ...891..134X}. Fitting for the monopole component allows us to constrain the upper limits of $T\rightarrow P$ leakage. The results of these fits are shown in~Figure~\ref{fig:ItoP_Moon}. Averaging over single detectors, a 95\% upper limit of (1.7$\pm$0.1)$\times$10$^{-3}$ is obtained. Using pair-differenced detectors, the 95\% upper limit on $T\rightarrow P$ leakage drops to (2.7$\pm$0.3)$\times$10$^{-4}$. Uncertainties on the upper limits are estimated from scatter among the individual measurements. While we should be cautious in interpreting these estimates because of the intrinsic Moon polarization signal, these estimates are comparable with the estimates obtained from Jupiter observations. Detectors on the central module 4 and adjacent modules 2 and 5 have the smallest upper limits. This is being further studied.

\begin{figure}
    \centering
    \includegraphics[width=1.0\linewidth]{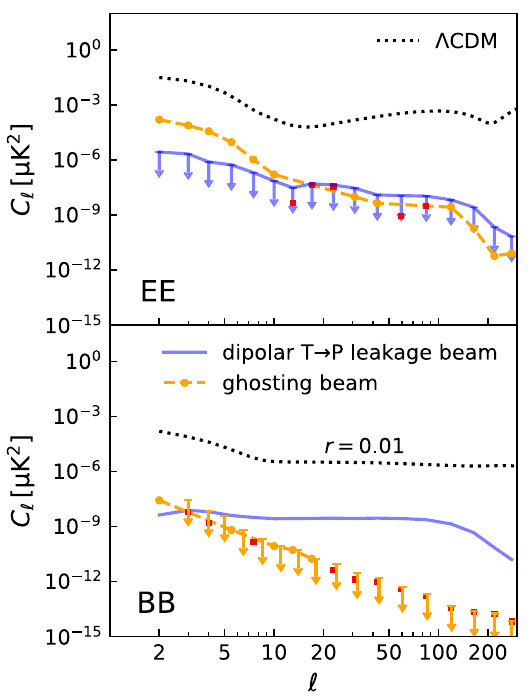}
    \caption{\textbf{
    Simulations of beam systematics induced by internal reflections and dipolar $T\rightarrow P$ leakage. } Modeled systematic (only) beams for dipolar $T\rightarrow P$ leakage and internal reflections (i.e., ghosting) are convolved with simulated sky temperature and linear polarization maps using \texttt{pisco} \citep{Fluxa20}. The simulated maps are based on \textit{Planck} CMB realizations with B-mode amplitude set to zero. The resulting power spectra are plotted and represent the expected systematic bias to the EE (top) and BB (bottom) power spectra. The curves with downward pointing arrows represent the 95\% confidence level upper limit for systematics that are not detected given the sample variance in the simulations. The red squares indicate negative values in the orange curves. These spectra are equivalent to the difference between the auto spectra from maps with systematics and without systematics. The expected impact of these systematics that are not accounted for in our cosmological beam model are at the subpercent level compared to the cosmological signals (dashed lines, BB corresponds to the primordial B-mode power spectrum for $r=0.01$)~\citep{2020A&A...641A...6P}.  }
    \label{fig:ItoP_impact}
\end{figure}

\vskip 5.8mm plus 1mm minus 1mm
\vskip1sp
\section{Impact of Beam Systematics}
\label{sec:systematics}
\vskip1sp

With the help of simulations and beam modeling we estimate the impact of our two major sources of beam systematics, i.e., due to the internal reflections (ghosting) and the dipolar $T\rightarrow P$ leakage, on our ability to unambiguously measure the CMB polarization power spectra. For modeling the internal reflection, the web structure of the field stop behind and in between the feedhorns is projected on to the detector coordinates. The ghosting beam is modeled using a combined reflectance (of the field stop and the suspected optical element near the cold stop) and location of the detector with respect to the center of the focal plane, assuming plane-mirror reflection along the optical axis. This model is then convolved with the main beam and the Moon
to fit the ghosting map of each detector as shown in Figure \ref{fig:ghosting}. The resultant best fit map, after deconvolving the Moon, serves as the model of the internal reflection or the ghosting beam. 
For constructing the dipolar $T\rightarrow P$ leakage beam model, we use the best-fit amplitudes of the dipole component of the Gauss–Hermite basis obtained from fitting to the measured beams.

CMB map realizations based on the \textit{Planck} best-fit parameters \citep{2020A&A...641A...6P} are used as simulation input.
The B-mode of these simulations is set to zero to show any E/B mixing effect clearly.
The impact of the systematics is assessed by comparing the EE/BB autocorrelation spectra before and after the inclusion of systematics.
For the ghosting beam and dipolar $T\rightarrow P$ leakage simulations, the corresponding beams are convolved with the polarization and temperature component of CMB simulations respectively, using a pixelized beam convolutor \texttt{pisco}~\citep{Fluxa20}. 
The resulting power spectra are shown in Figure~\ref{fig:ItoP_impact}; in both cases, the systematics errors are subdominant compared to the cosmological signal.

\vskip 5.8mm plus 1mm minus 1mm
\vskip1sp
\section{Summary}
\label{sec:summary}
\vskip1sp
In this work, we characterized the optical performance of the first CLASS $90\,\mathrm{GHz}$ telescope. We described measurements of the far-field $90\,\mathrm{GHz}$ telescope beam using observations of Jupiter. The effective instrument beam FWHM and solid angle integrated to 4$^{\circ}$ are $0.620\pm0.003$\dg{} and $138.7\pm0.6$(stats.)$\pm1.1$(sys.)$\mu{\rm sr}$, respectively.

We also presented the Moon-based telescope pointing calibration. The pointing model is updated periodically to account for temporal changes in beam pointing and optical alignment. The standard deviations of the pointing error with respect to the model as inferred from recurring Moon scans across the observing epoch are 1.3$^{\prime}$, 2.1$^{\prime}$, and 2.0$^{\prime}$, respectively, in azimuth, elevation, and boresight rotation angle. This corresponds to a combined pointing uncertainty of $2.5'$ ($\sim$7$\%$ of the beam FWHM), implying a $\sim$3\% broadening of the cosmological survey beam. The average differential pointing between co-located $\pm$45\dg{} detectors is 1.2$^{\prime}$ ($\sim$3$\%$ of the beam FWHM) with a scatter of $\pm$0.7$^{\prime}$ across the focal plane.

We measured the Tau A polarization angle to be 149.6\dg{} with a statistical uncertainty of 0.2\dg{}. This is consistent with the range of other measurements near $90\,\mathrm{GHz}$ when accounting for systematic uncertainties of the various measurements and establishes calibration of the instrument polarization angle. 

Far-sidelobes characterized using per-detector centered Moon maps in intensity and polarization showed sub-percent level ghosting. By comparing polarization and intensity maps generated from dedicated Jupiter scans, we estimated from averaging over single detectors that $1.7\times$10$^{-3}$ of the intensity signal leaks into polarization as a monopole component with an additional dipolar leakage at the level of $4.3\times$10$^{-3}$. Averaging over pair-differenced detectors, we place a 95\% confidence upper limit of $5.5\times$10$^{-4}$ on the $T\rightarrow P$ leakage fraction. We did not find any evidence of quadrupolar $T\rightarrow P$ leakage. 

The impact of the measured beam systematics associated with the internal reflections and the dipolar $T\rightarrow P$ leakage were assessed through modeling and simulations. We found that these systematics, unaccounted for in the cosmological analysis, would lead to sub-percent level bias in the inferred CMB polarization power spectra. 

\vskip 5.8mm plus 1mm minus 1mm
\vskip1sp
\section*{Acknowledgments}
\vskip4pt

We acknowledge the National Science Foundation Division of Astronomical Sciences for their support of CLASS under Grant Numbers 0959349, 1429236, 1636634, 1654494, 2034400, and 2109311. We thank Johns Hopkins University President R. Daniels and the Deans of the Kreiger School of Arts and Sciences for their steadfast support of CLASS. We further acknowledge the very generous support of Jim and Heather Murren (JHU A\&S ’88), Matthew Polk (JHU A\&S Physics BS ’71), David Nicholson, and Michael Bloomberg (JHU Engineering ’64). The CLASS project employs detector technology developed in collaboration between JHU and Goddard Space Flight Center under several NASA grants. Detector development work at JHU was funded by NASA cooperative agreement 80NSSC19M0005. CLASS is located in the Parque Astron\'omico Atacama in northern Chile under the auspices of the Agencia Nacional de Investigaci\'on y Desarrollo (ANID). 

We acknowledge scientific and engineering contributions from Max Abitbol, Fletcher Boone, Jay Chervenak, Lance Corbett, David Carcamo, Manwei Chan, Kevin~L. Denis, Mauricio D\'iaz, Rolando D\"{u}nner, Dominik Gothe, Ted Grunberg, Saianeesh Haridas, Connor Henley, Gene Hilton, Johannes Hubmayr, Ben Keller, Lindsay Lowry, Nick Mehrle, Nathan Miller, Grace Mumby, Keisuke Osumi, Gonzalo Palma, Diva Parekh, Isu Ravi, Carl~D. Reintsema, Daniel Swartz, Bingjie Wang, Qinan Wang, Emily Wagner, Tiffany Wei, Zi\'ang Yan, Lingzhen Zeng, and Zhuo Zhang. For essential logistical support, we thank Jill Hanson, William Deysher, Joseph Zolenas, LaVera Jackson, Miguel Angel D\'iaz, Mar\'ia Jos\'e Amaral, and Chantal Boisvert. We acknowledge productive collaboration with Dean Carpenter and the JHU Physical Sciences Machine Shop team. I.L.P. gratefully acknowledges support from the Horizon Postdoctoral Fellowship. 

S.D. is supported by an appointment to the NASA Postdoctoral Program at the NASA Goddard Space Flight Center, administered by Oak Ridge Associated Universities under contract with NASA. K.H. is supported by NASA under award number 80GSFC21M0002. R.R. acknowledges partial support from CATA, BASAL grant AFB-170002, and CONICYT-FONDECYT through grant 1181620. Z.X. is supported by the Gordon and Betty Moore Foundation through grant GBMF5215 to the Massachusetts Institute of Technology.

\software{We used various software packages for this work including \texttt{IPython} \citep{ipython}, \texttt{numpy} \citep{numpy}, \texttt{scipy} \citep{2020NatMe..17..261V}, \texttt{matplotlib} \citep{matplotlib}, \texttt{healpy} \citep{healpix, healpy}, 
\texttt{PyEphem} \citep{pyephem}, and \texttt{Astropy} \citep{astropy1, astropy2}}

\bibliography{references}


\appendix

\begin{deluxetable}{@{\extracolsep{2pt}}ccccccc}
\tablenum{A1}
\tablecaption{Fore-optics parameters.}
\label{tb:fore_optics}
\tablehead{
\colhead{} & \multicolumn{2}{c}{Center} & \multicolumn{2}{c}{focus 1} & \multicolumn{2}{c}{focus 2} \\
\cline{2-3} \cline{4-5} \cline{6-7}
\colhead{Element} & \colhead{x (cm)} & \colhead{y (cm)} & \colhead{x (cm)} & \colhead{y (cm)} &
\colhead{x (cm)} & \colhead{y (cm)}
}
\startdata
VPM &  131.8 & 102.7 & & & & \\
Primary & $-43.5$ & 148.8 &  130.8 & $-9.1$ & 362.9 & 43.5 \\
Secondary & 118.5 & 0 &  $-87.2$ & 193.7 &  $-193.4$ & $-3.9$ \\
\enddata 
\tablecomments{The coordinates are defined with the aperture stop placed at the origin and the $x$-axis aligned towards the center of the secondary mirror.}
\end{deluxetable}

\begin{deluxetable*}{@{\extracolsep{2pt}}ccccccccccc}
\tablenum{A2}
\tablecaption{Reimaging optics parameters.}
\label{tb:reim_optics}
\tablehead{
\colhead{} & \colhead{} & \colhead{} & \colhead{} & \colhead{} & \colhead{index of} & \multicolumn{2}{c}{surface 1} & \multicolumn{2}{c}{surface 2} \\
\cline{7-8} \cline{9-10}
\colhead{Element} & \colhead{material} & \colhead{label} & \colhead{dist (cm)} &\colhead{thickness (cm)} & \colhead{refraction} & \colhead{R (cm)} & \colhead{$\kappa$} & \colhead{R (cm)} & \colhead{$\kappa$}
}
\startdata
4K filter & PTFE & A & 1.5 & 1.0 & 1.46 & & & & \\
60K filter 1 & PTFE & B & 7.92 & 1.0 & 1.46 & & & & \\
60K filter 2 & PTFE & C & 10.67 & 1.0 & 1.453 & & & & \\
Window & UHMWPE & D & 29.37 & 0.48 & 1.52 & & & & \\
4K lens & HDPE & E & 6.00 & 5.05 & 1.564 & 162.37 &  $-0.018$ & $-55.05$ & $-2.170$ \\
1K filter & Nylon & F & 57.36 & 1.0 & 1.72 & & & & \\
1K lens & HDPE & G & 59.36 & 6.10 & 1.564 & 78.37 & $-10.460$ & $-56.45$ & $-2.487$ \\
Focal Plane & & H & 90.41 & & & & \\
\enddata
\tablecomments{Distance labels are as indicated in Figure~\ref{fig:Wband_fore_optics}. All surfaces are modeled as planar unless otherwise specified.}
\end{deluxetable*}

\section{Optical Design}
\label{sec:opt_des}

The optical design of the CLASS $90\,\mathrm{GHz}$ telescope follows from the $40\,\mathrm{GHz}$ telescope design described in~\cite{eime12}. Here, we briefly summarize the design for the $90\,\mathrm{GHz}$ telescope and present the main design parameters. The design can be divided into a \emph{fore-optics} section, including all elements between the cold aperture stop and the sky, and a \emph{reimaging} section, which ultimately forms a focal plane using two lenses. 

\begin{figure}
    \centering
    \includegraphics[width=1\linewidth]{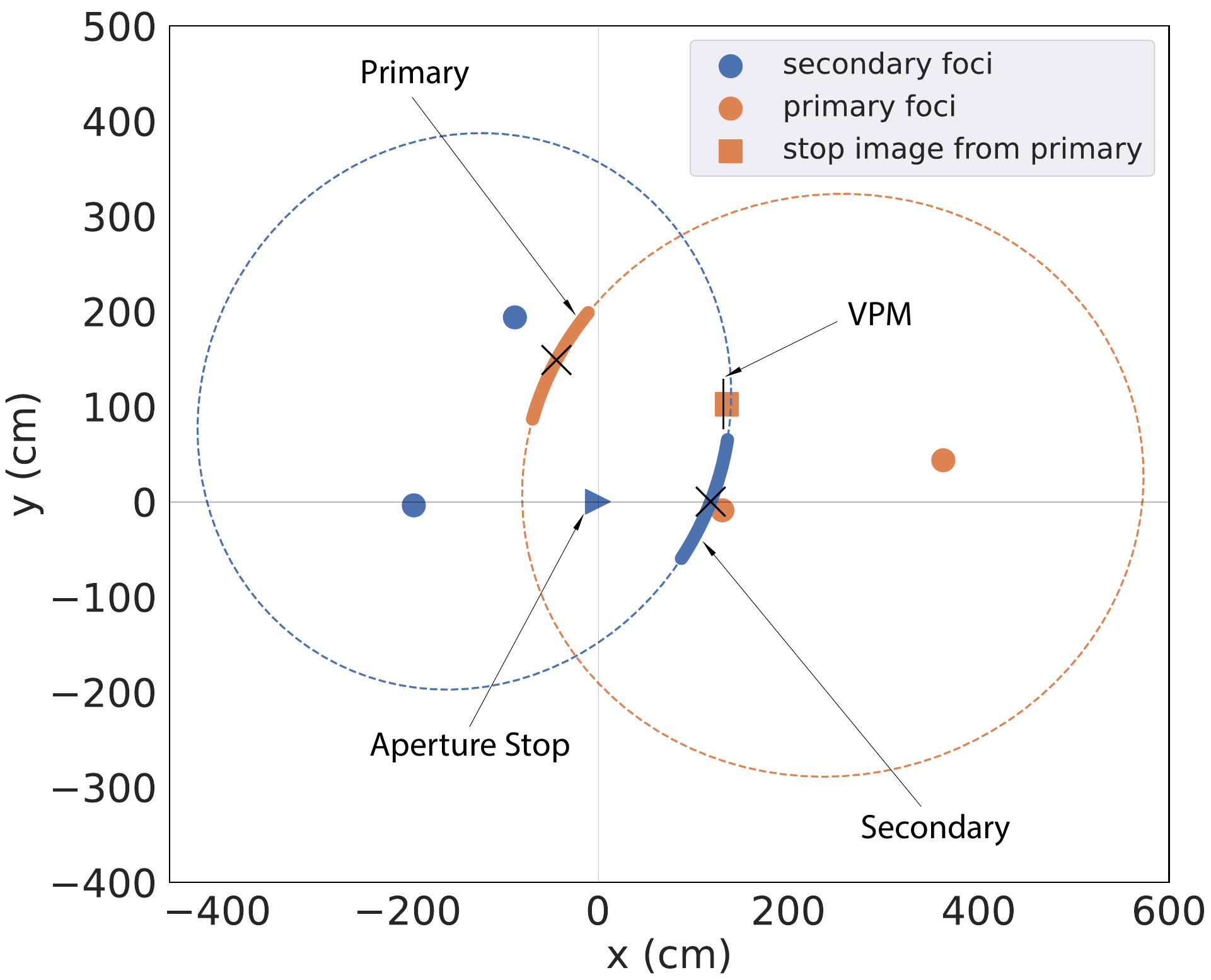}
    \caption{{\textbf{Fore-optics design concept.}} Each mirror is a section of an off-axis ellipsoid with foci indicated by \markertwo\,\ (for primary), \markerone\,\ (for secondary), and mirror center locations indicated by `$\times$' marks, and specified in Table~\ref{tb:fore_optics}. The final image of the stop is indicated by \markerfour. This fore-optics system places the entrance pupil on the VPM. } 
    \label{fig:Wband_fore_optics}
\end{figure}

The fore-optics consist of two mirrors and the front-end modulator, i.e., the VPM. For the purpose of the geometric design, the VPM is modeled as a folding flat. The mirrors' shapes and positions are optimized to form an image of the aperture stop on the VPM, thus placing the entrance pupil of the telescope at the front-end polarization modulator. Figure~\ref{fig:Wband_fore_optics} shows the fore-optics mirror shapes. Both mirrors are off-axis sections of ellipsoids generated by rotating the parent ellipse shape about their major axis. The mirrors are defined by the focal points of the parent ellipse and a point at the center of each mirror, see Table~\ref{tb:fore_optics}. 

\begin{figure*}
    \centering
    \includegraphics[width=0.9\linewidth]{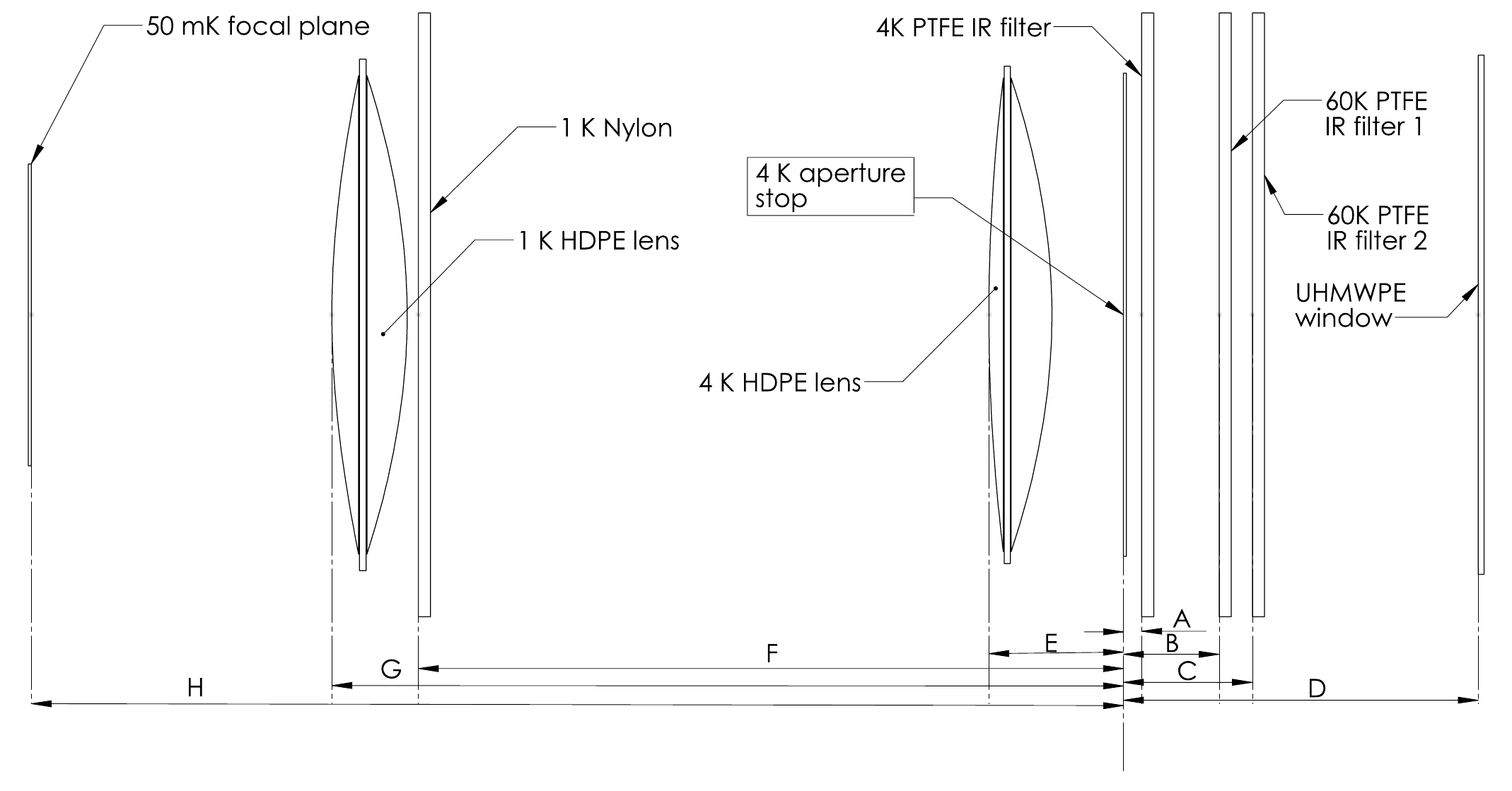}
    \caption{{\textbf{Reimaging optics}}. The reimaging optics use two lenses to form the image on the focal plane. The location of each element, referenced to the aperture stop, is specified in Table~\ref{tb:reim_optics}. Additional metal-mesh and foam infrared filters are used in the receiver, but these do not impact the optical model.} 
    \label{fig:Wband_reim_optics}
\end{figure*}

The reimaging optics consist of two high-density polyethylene (HDPE) lenses. Flat dielectric IR blocking filters are included in the optical model as they can impact the optimal axial positions of the other elements. The lenses and filters are held within the cryogenic receiver with a UHMWPE vacuum window, which is also included in the model. Each lens surface is a simple conic section defined as:
\begin{equation} 
    z = \frac{cr^{2}}{1+\sqrt{1-(1+\kappa) c^{2} r^{2}}},
\end{equation}
\noindent where, $z$ is the sag of the surface, $r$ is the radial polar coordinate, c is the curvature (or inverse of the radius of curvature), and $\kappa$ is the conic constant. Table~\ref{tb:reim_optics} describes the shapes and positions of the lenses and IR blocking filters, also see Figure~\ref{fig:Wband_reim_optics}. All distance and shape parameters describe the optical system in its cold operating condition. For machining purposes, the lenses are assumed to experience an integrated expansion of 2\% in length once at room temperature. In addition to thickness, the radius of curvature of each surface also experiences this same expansion; the conic constants are assumed fixed. The receiver design encompasses the opto-thermo-mechanical strategy to ensure proper positioning of the optical elements when at the operating temperature. 

The surface of each lens included a small sacrificial layer, beyond the described thicknesses above, into which an array of sub-wavelength holes are patterned. The hole diameter, depth, and the spacing between holes determine the effective index of this metamaterial layer~\citep{1262850}. These parameters were optimized so that the metamaterial layers function as AR coatings. A similar AR coating was separately optimized for the 1K nylon filter. The AR coatings for the polytetrafluroethylene (PTFE) filters and cryostat window were formed by heat-pressing porous PTFE sheets onto each side of the UHMWPE window and PTFE filters with thin low-density polyethylene bonding layers. The thicknesses of the AR layers were optimized for maximum in-band transmission. For details on the AR coating design and implementation, see~\cite{Dahal:thesis}.

The design presented here, similar to the $40\,\mathrm{GHz}$ telescope design~\citep{eime12}, was modeled using the Zemax OpticStudio optical design software~\footnote{https://www.zemax.com/pages/opticstudio} and has a simulated performance exceeding a Strehl ratio of 0.98 over the full telescope field of view. 

\end{document}